\documentclass[a4paper, fleqn, usenatbib, useAMS]{mnras}

\usepackage{graphicx} 
\usepackage{amsmath, amssymb} 
\usepackage{newtxtext, newtxmath} 
\usepackage[T1]{fontenc}
\usepackage{ae, aecompl}
\usepackage{xcolor}

\title[Radio masers on WX UMa]{Radio masers on WX UMa: hints of a Neptune-sized planet, or magnetospheric reconnection?}

\author[Kavanagh et al.]{Robert D. Kavanagh$^{1, 2}$\thanks{Contact e-mail: \href{kavanagh@strw.leidenuniv.nl}{kavanagh@strw.leidenuniv.nl}}, Aline A. Vidotto$^{1, 2}$, Harish K. Vedantham$^{3,4}$, \newauthor Moira M. Jardine$^{5}$, Joe R. Callingham$^{1,3}$, Julien Morin$^{6}$ \\ 
$^1$Leiden Observatory, Leiden University, PO Box 9513, 2300 RA, Leiden, The Netherlands \\
$^2$School of Physics, Trinity College Dublin, The University of Dublin, Dublin 2, Ireland \\
$^3$ASTRON, The Netherlands Institute for Radio Astronomy, Oude Hogeveensedijk 4, 7991PD, Dwingeloo, The Netherlands \\
$^4$Kapteyn Astronomical Institute, University of Groningen, Landleven 12, 9747 AD Groningen, The Netherlands \\
$^5$SUPA, School of Physics and Astronomy, University of St Andrews, St Andrews KY16 9SS, UK \\
$^6$Laboratoire Univers et Particules de Montpellier (LUPM), Universit\'e de Montpellier, CNRS, 34095 Montpellier, France}

\date{Last updated ...; in original form ...}

\pubyear{2022}

\graphicspath{{figs/}}

\begin{document}
\label{firstpage}
\pagerange{\pageref{firstpage}--\pageref{lastpage}}
\maketitle


\begin{abstract}
The nearby M dwarf WX UMa has recently been detected at radio wavelengths with LOFAR. The combination of its observed brightness temperature and circular polarisation fraction suggests that the emission is generated via the electron-cyclotron maser instability. Two distinct mechanisms have been proposed to power such emission from low-mass stars: either a sub-Alfv\'enic interaction between the stellar magnetic field and an orbiting planet, or reconnection at the edge of the stellar magnetosphere. In this paper, we investigate the feasibility of both mechanisms, utilising the information about the star's surrounding plasma environment obtained from modelling its stellar wind. Using this information, we show that a Neptune-sized exoplanet with a magnetic field strength of 10 -- 100~G orbiting at $\sim0.034$~au can accurately reproduce the observed radio emission from the star, with corresponding orbital periods of 7.4~days. Due to the stellar inclination, a planet in an equatorial orbit is unlikely to transit the star. While such a planet could induce radial velocity semi-amplitudes from 7 to 396~m~s$^{-1}$, it is unlikely that this signal could be detected with current techniques due to the activity of the host star. The application of our planet-induced radio emission model here illustrates its exciting potential as a new tool for identifying planet-hosting candidates from long-term radio monitoring. We also develop a model to investigate the reconnection-powered emission scenario. While this approach produces less favourable results than the planet-induced scenario, it nevertheless serves as a potential alternative emission mechanism which is worth exploring further.
\end{abstract}



\section{Introduction}
\label{sec:intro}

Recent observations with the \textit{LOw Frequency ARray} \citep[LOFAR,][]{vanhaarlem13} have begun to shed light on the coherent radio emission mechanisms at play in the coronae of M~dwarfs \citep{vedantham20, davis21, callingham21b}. Many of these observations could be due to interactions between the star's magnetic field and a planet orbiting in the sub-Alfv\'enic region of the stellar wind, which can extend out to tens of stellar radii in the case of M~dwarfs \citep{davis21, kavanagh21}. Within 0.1~au, M~dwarfs are expected to host numerous rocky exoplanets \citep{burn21}, the same region where the habitable zones of these stars are thought to lie \citep{kopparapu13}. Confirmation of such interactions at radio wavelengths would be a game-changer in the world of exoplanet detection, as these types of exoplanets remain largely undiscovered by other techniques. They could also provide a new avenue to probe both the magnetospheres of exoplanets, as well as the stellar wind environment of the host star \citep{kavanagh21}.

One such M~dwarf that was recently detected by LOFAR as part of the LOFAR Two-metre Sky Survey \citep[LoTSS,][]{shimwell17} is the nearby active star WX Ursae Majoris (hereafter WX~UMa) \citep{callingham21b}. It is a fast rotator, and exhibits a strong and predominantly dipolar magnetic field, with an estimated average unsigned strength of $\sim1$~kG derived from Stokes V observations \citep{morin10}. Additionally, Zeeman broadening measurements of the star suggest that the surfaced-averaged field strength could be up to $7$~kG, the strongest field measured for a cool main-sequence star \citep{shulyak17}. The relevant physical parameters of WX~UMa are listed in Table~\ref{tab:star}. 

Between 2014 and 2016, WX~UMa was detected nearly continuously in the radio over three 8-hour intervals \citep{davis21, callingham21b}. The observed emission exhibits a high degree of circular polarisation ($\gtrsim70\%$), with a peak flux density of $\sim1.2$~mJy. The combined brightness temperature and high degree of circular polarisation of the emission implies that it is generated via a coherent emission mechanism. Two types of processes can produce coherent radio emission: plasma and cyclotron emission \citep{dulk85}. Plasma emission is powered by the conversion of turbulent Langmuir wave energy into electromagnetic energy. Such a scenario can occur when hot plasma is injected into a cooler and denser one, such as during the flaring of coronal loops \citep{zaitsev83, stepanov01, vedantham21}. However, \citet{callingham21b} illustrated that the emission from WX~UMa cannot be explained by plasma emission, due to the combination of its observed brightness temperature and circular polarisation fraction. Therefore, cyclotron emission is likely to be the mechanism producing the observed radio emission.

Cyclotron emission, or electron-cyclotron maser instability (ECMI) emission, occurs in rarefied, strongly magnetised plasmas \citep{dulk85}. The basic principle in a stellar/planetary context is as follows: electrons are accelerated along magnetic field lines, travelling towards regions of higher magnetic field strengths. This introduces a population of high-velocity electrons into the plasma relative to the already-existing thermal distribution. As the field strength increases, the field lines converge, and the accelerated electrons are reflected due to a magnetic mirroring effect, provided that their pitch angle (the angle between their velocity vector and the magnetic field) is large enough. This produces a so-called `loss cone' or `horseshoe' distribution, referring to the shape of the velocity space that the electrons which power the maser occupy \citep{treumann06}. These conditions have been identified as those necessary to drive the maser. If electrons are continuously accelerated, the maser will continue to be powered. The first star discovered to exhibit emission consistent with ECMI was the chemically peculiar star CU Vir \citep{trigilio00}.

Two sources of the energy required to power ECMI have been identified for stars. The first of these is analogous of the sub-Alfvénic interactions between Jupiter and Io \citep{neubauer80, hess08, saur13}. In this scenario, Io perturbs Jupiter's magnetic field in a sub-Alfvénic orbit, producing Alfvén waves which travel back towards Jupiter. The mechanical energy carried by the waves is then thought to subsequently dissipate, accelerating electrons and powering ECMI. For close-in exoplanetary systems, a similar process is thought to occur, where the roles of Jupiter and Io are taken by the star and planet respectively. Such interactions are expected to produce bright radio emission, at both MHz \citep{hess11, turnpenney18, kavanagh21} and GHz frequencies \citep{leto17, pereztorres21}. The current sheet regions of the plasma environments surrounding stars \citep{linsky92, trigilio04, nichols12, owocki22} and planets \citep{cowley01} have also been identified as suitable acceleration sites for electrons in powering ECMI. In this scenario, magnetic reconnection is thought to provide electrons with energy at the edge of the magnetosphere, accelerating them to higher latitudes.

In this paper, we explore the feasibility of these two scenarios in reproducing the observed radio emission from WX~UMa, utilising the plasma environment obtained from modelling its stellar wind environment. For the planet-induced scenario, we expand upon the model presented by \citet{kavanagh21}, accounting for the beaming and polarisation of the generated emission, as well as the stellar rotational and planetary orbital motions for the first time in the literature. We then apply our updated planet-induced radio emission model to WX~UMa, and investigate what planetary and orbital properties best-reproduce the radio observed emission of WX~UMa. We also develop a model to investigate the reconnection-powered scenario.

\begin{table}
\caption{Stellar parameters of WX UMa used in this work.}
\label{tab:star}
\centering
\begin{tabular}{lc}
\hline
Stellar parameter & Value \\
\hline
Mass ($M_\star$) $^1$ & 0.095~$M_{\sun}$ \\
Radius ($R_\star$) $^2$ & 0.12~$R_{\sun}$ \\
Unsigned average large-scale magnetic field strength $^2$ & 1~kG \\
Rotation period ($P_\star$) $^2$ & 0.78~days \\
Inclination ($i_\star$) $^2$ & 40$\degr$ \\
Distance ($d$) $^3$ & 4.9~pc \\
\hline
\multicolumn{2}{p{0.8\columnwidth}}{1: \cite{newton17}; 2: \cite{morin10}; 3: \cite{gaia18}} \\
\hline
\end{tabular}
\end{table}


\section{Stellar wind environment of WX UMa}
\label{sec:wind model}

To model the stellar wind of WX UMa, we use the Alfv\'en wave-driven \textsc{AWSoM} model \citep{vanderholst14} implemented in the 3D magnetohydrodynamics (MHD) code \textsc{BATS-R-US} \citep{powell99}. In this model, Alfv\'en waves propagate outwards from the base of the chromosphere along the stellar magnetic field lines. As they propagate, the waves are partially reflected. The interaction of the outward-propagating and reflected waves produces a so-called `turbulent cascade', which dissipates the mechanical wave energy into thermal energy, heating the corona and driving the stellar wind outflow \citep{chandran11}.

\textsc{BATS-R-US} iteratively solves the ideal set of MHD equations in the stellar co-rotating frame on a three dimensional grid for a set of inputs, providing us with the density, magnetic field, velocity, pressure, current density, and Alfv\'en wave energy density of the stellar wind plasma in the observer's reference frame \citep{cohen11, vidotto12, garraffo17, alvaradogomez19, kavanagh21, evensberget22}. For our stellar wind model of WX UMa, we use a spherical grid that extends from the base of the chromosphere to 100 times the stellar radius. Our grid also includes regions of enhanced resolution, in order to mitigate the effects of numerical dissipation. The total number of cells in our grid is around 8,000,000. Once the mass-loss rate of the star changes by less than 10 percent between iterations, we take this to be the steady-state solution for our set of input values.

\begin{figure}
\includegraphics[width = \columnwidth]{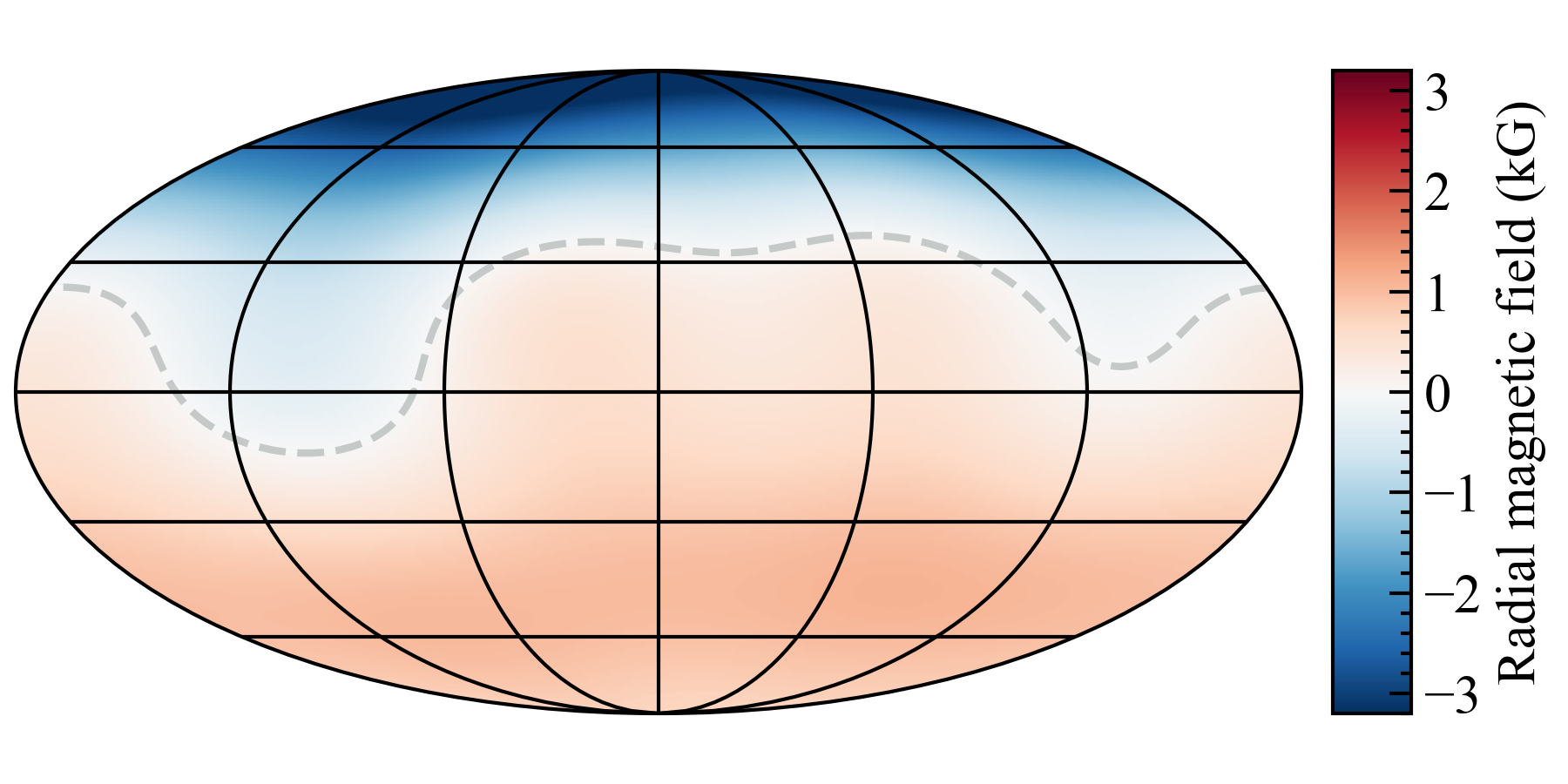}
\caption{Radial surface magnetic field map of WX UMa in 2006, reconstructed by \citet{morin10}. We implement this at the inner boundary in our stellar wind model. The dashed grey line shows where the polarity is neutral ($B_r = 0$). We refer to the visible hemisphere (the pole with a negative polarity) as the Northern hemisphere.}
\label{fig:zdi map}
\end{figure}

The main inputs for the AWSoM model are the mass, radius, and rotation period of the star, as well as the stellar surface magnetic field. For the magnetic field we use the map of the star in 2006, which was reconstructed by \citet{morin10} using the Zeeman-Doppler imaging method. The radial component of the map is implemented at the inner boundary of our wind model, which is shown in Figure~\ref{fig:zdi map}. Another key parameter in the model is the Alfv\'en wave flux-to-magnetic field ratio $S_A/B$, which affects the amount of mass lost via the stellar wind \citep[see][]{borosaikia20, kavanagh21, ofionnagain21}. This is a free parameter in our model. For WX UMa, we adopt an Alfv\'en wave flux-to-magnetic field ratio of $S_A/B = 1\times10^5$~erg~s$^{-1}$~cm$^{-2}$~G$^{-1}$. This produces a stellar wind with a mass-loss rate of $2.1\times10^{-14}~M_{\sun}~\textrm{yr}^{-1}$. With an X-ray luminosity of $3.1\times10^{27}$~erg~s$^{-1}$ \citep{schmitt04}, the mass-loss rate we obtain for WX~UMa is in agreement with emerging trends between the mass-loss rates and surface X-ray fluxes of low-mass stars \citep[see][]{jardine19, wood21, vidotto21}. Note that for the remaining inputs in the model relating to the Alfv\'en wave physics, we adopt those presented in \citet{kavanagh21}. In Appendix~\ref{sec:plasma properties}, we show the density, electron temperature, and magnetic field strength profiles of the stellar wind of WX~UMa.

\begin{figure*}
\includegraphics[width = 0.6\textwidth]{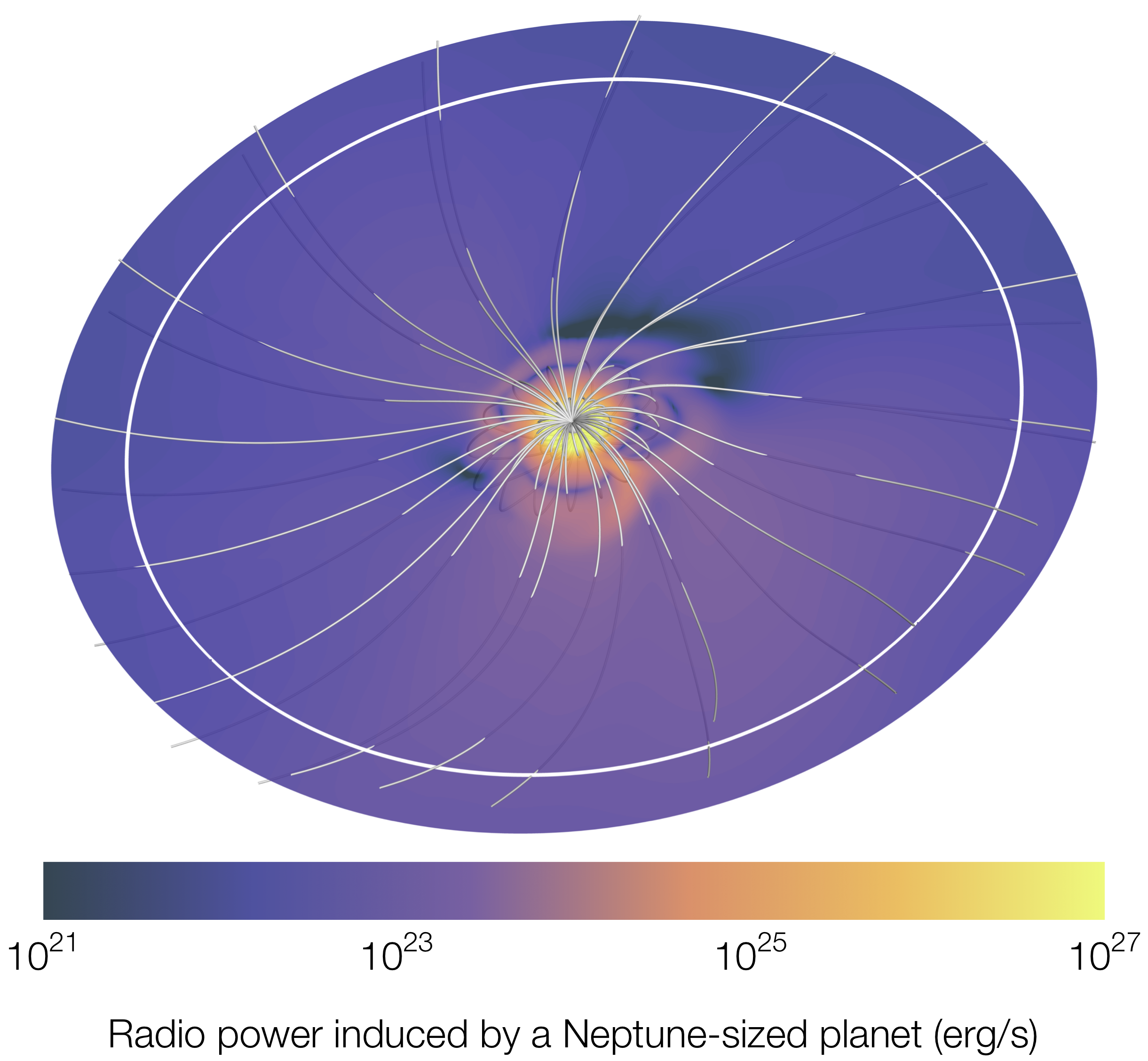}
\caption{Simulated stellar wind environment of WX UMa out to a distance of 70~$R_\star$ ($\sim0.039$~au). The equatorial plane is coloured with the radio power that can be induced in the star via Alfv\'en waves generated through a sub-Alfvénic interaction with a Neptune-sized planet with a 10~G magnetic field at that location. Around 1\% of the Alfv\'en wave energy is expected to be converted into radio emission. In the case of WX~UMa, sub-Alfvénic interactions can occur out to $\sim80~R_\star$ in the equatorial plane. The white circle shows the orbital distance (0.034~au) we identify where a Neptune-sized exoplanet could orbit and induce emission at 144~MHz, comparable to that recently detected with LOFAR (see Section~\ref{sec:best fit}). The grey lines show the large-scale magnetic field of the star.}
\label{fig:wind}
\end{figure*}

With a mass-loss rate of $2.1\times10^{-14}~M_{\sun}~\textrm{yr}^{-1}$, we find that the stellar wind of WX~UMa is sub-Alfv\'enic out to $\sim60$--80~$R_\star$, in the reference frame of a planet orbiting in the equatorial plane. We also find that the closed-field region of the star's magnetic field extends out to $\sim40$~$R_\star$ in the current sheet (where the radial magnetic field is zero). A 3D view of the stellar wind environment is shown in Figure~\ref{fig:wind}. The location of the Alfv\'en surface of the stellar wind in the equatorial plane is of particular interest, as inside this region, an orbiting planet can induce the generation of radio emission along the magnetic field line connecting the planet and star. In the reference frame of an orbiting planet, the sub-Alfv\'enic region is defined as where the relative velocity between the stellar wind and the planet ($\Delta u$) is less than the local Alfv\'en velocity:
\begin{equation}
\Delta u < u_\textrm{A} = \frac{B_\textrm{w}}{\sqrt{4 \pi \rho_\textrm{w}}} ,
\label{eq:alfven velocity}
\end{equation}
where $B_\textrm{w}$ and $\rho_\textrm{w}$ are the magnetic field strength and density of the stellar wind at the position of the planet.

The location of the Alfv\'en surface in the current sheet of the stellar wind plasma is also of interest in the context of radio emission, in that it is thought to be a point of significant acceleration for electrons at the edge of planetary and stellar magnetospheres, powered via magnetic reconnection \citep{linsky92, trigilio04, cowley01, nichols12, owocki22}. In the following Sections, we explore the feasibility of both the planet-induced and reconnection scenarios in producing the observed emission of WX~UMa at 144~MHz, utilising the information about the plasma environment obtained from our modelling of the stellar wind.

\section{Emission induced by an orbiting planet}
\label{sec:planet radio emission}

If there is a planet orbiting inside the Alfv\'en surface of WX~UMa (i.e. at an orbital distance of $\lesssim80~R_\star$), it can perturb the magnetic field of the star and produce Alfv\'en waves \citep{neubauer80}. These waves can travel back to the star along the field lines connecting to the planet, producing radio emission via ECMI \citep{saur13, turnpenney18}. In \citet{kavanagh21}, we developed a model to compute the emission generated in this type of interaction, which accounts for the realistic magnetic field structure and plasma environment of the star obtained from stellar wind simulations such as that shown in Figure~\ref{fig:wind}. Here, we expand upon this model further, accounting for both the beaming and polarisation of the generated emission, as well as the respective stellar rotational and orbital motions of the potential planet.

\begin{figure}
\includegraphics[width = \columnwidth]{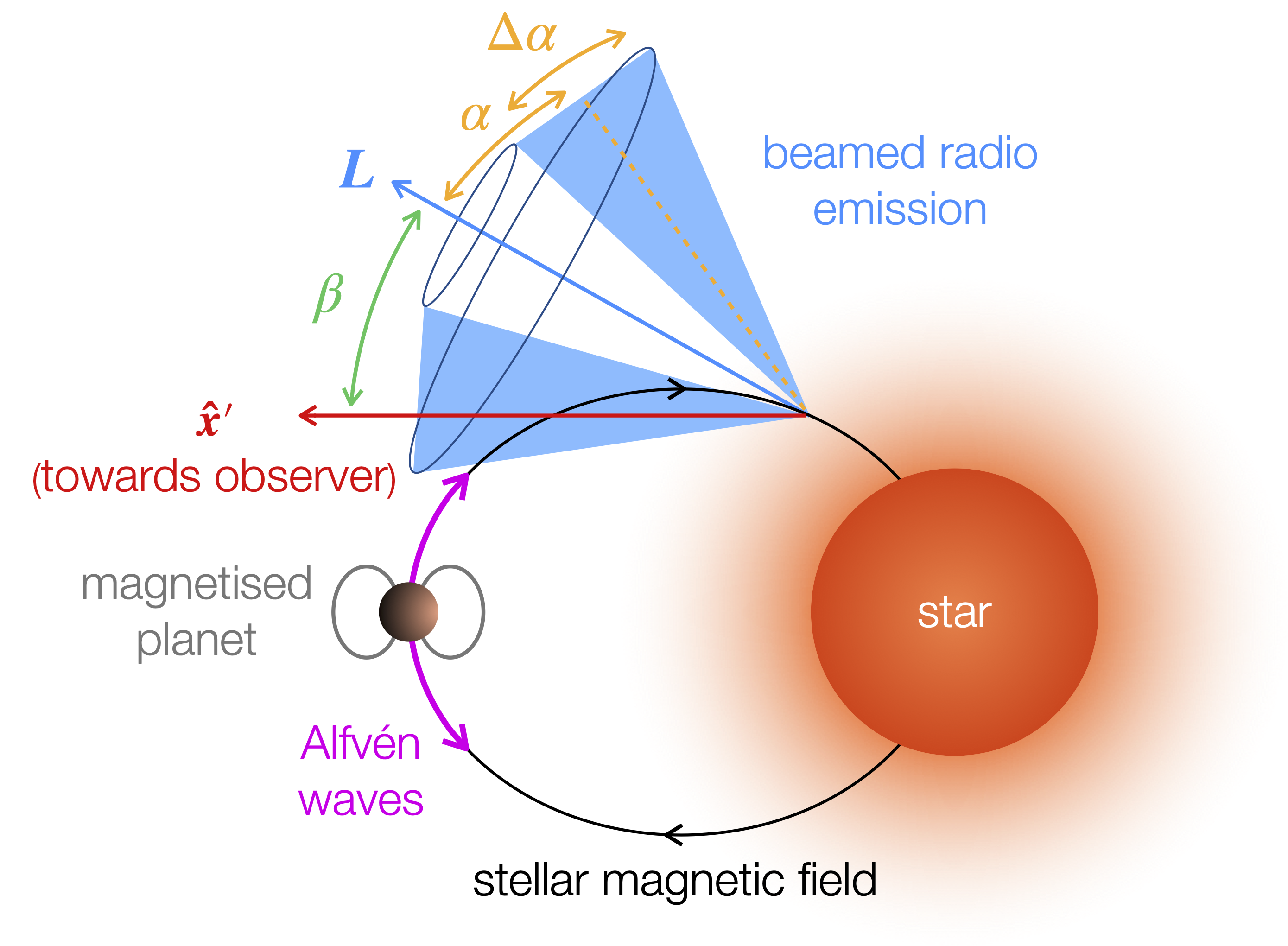}
\caption{Sketch of the emission cone geometry for the planet-induced scenario. If a planet orbits its host star with a sub-Alfv\'enic velocity (Equation~\ref{eq:alfven velocity}), it can produce Alfv\'en waves which carry energy back towards the star along the field line connecting the star and planet. This energy then dissipates near the surface, producing radio emission via ECMI. In the closed-field region of the star's magnetic field, the field line connects the planet back to the star in both hemispheres, producing emission in two locations. For clarity here we show emission generated in the Northern hemisphere only. If the planet orbits in the open-field region however, the generation of ECMI is limited to one hemisphere. The emission generated is beamed in a hollow cone, with an opening angle $\alpha$ and a thickness $\Delta\alpha$. The angle $\alpha$ is measured from the tangent to the magnetic field line $\boldsymbol{L}$ which points away from the stellar surface, in the opposite direction of the Alfvén wave propagation. The vector $\boldsymbol{\hat{x}}'$ points towards the observer, and $\beta$ is the angle between $\boldsymbol{L}$ and $\boldsymbol{\hat{x}}'$. If $\beta$ is in the range of $\alpha\pm\Delta\alpha$, the emission is seen by the observer.}
\label{fig:sketch cone}
\end{figure}

ECMI emission is beamed in a hollow cone \citep{dulk85}, which will only be seen by the observer if the beam points along the line of sight. We take the emission cone to have an opening angle $\alpha$ and thickness $\Delta \alpha$. The angle $\alpha$ is measured from the tangent $\boldsymbol{L}$ of the magnetic field that points away from the stellar surface, and the unit vector $\boldsymbol{\hat{x}}'$ points towards the observer. For WX~UMa, the visible (Northern) hemisphere exhibits a predominantly negative polarity (Figure~\ref{fig:zdi map}), and therefore $\boldsymbol{L} = - \boldsymbol{B}$, where $\boldsymbol{B}$ is the magnetic field. A sketch of the geometry described here is shown in Figure~\ref{fig:sketch cone}.

The angle formed between the emission cone and line of sight in the Northern hemisphere is given by
\begin{equation}
\cos \beta = \frac{\boldsymbol{L} \cdot \boldsymbol{\hat{x}}'}{B \hat{x}'} = \frac{- \boldsymbol{B} \cdot \boldsymbol{\hat{x}}'}{B \hat{x}'} = \frac{-{{B_x}'}}{B} ,
\label{eq:beta}
\end{equation}
where ${B_x}'$ is the magnetic field component at the emission site that points towards the observer. Note that in the Southern hemisphere, the tangent vector is $\boldsymbol{L} = \boldsymbol{B}$. In general, the radial magnetic field at the emitting point $B_r$ tells us the direction of the tangent vector. The emission is beamed towards the observer if the angle $\beta$ is in following range:
\begin{equation}
\alpha - \frac{\Delta \alpha}{2} < \beta < \alpha + \frac{\Delta \alpha}{2} .
\label{eq:beta condition}
\end{equation}
100\% circularly polarised ECMI emission is generated at either the fundamental or second harmonic of the local cyclotron frequency:
\begin{equation}
\nu_\textrm{c} = \frac{e B}{2 \pi m_e},
\label{eq:cyclotron frequency}
\end{equation}
where $e$ and $m_e$ are the electron charge and mass respectively. Note that for ECMI emission to be generated, the cyclotron frequency must exceed the local plasma frequency \citep{dulk85}:
\begin{equation}
\nu_\textrm{c} > \nu_\textrm{p} = \sqrt{\frac{e^2 n_e}{\pi m_e}},
\label{eq:plasma frequency}
\end{equation}
where $n_e$ is the electron number density.

The polarisation of ECMI emission depends on the magnetic polarity of the point on the field line where it is generated relative to the observer, as well as the magnetoionic mode. For LOFAR, the convention for emission in the \textit{o}-mode is that the Stokes V flux is observed with a positive sign for a positive polarity ($B_r > 0$), and negative for a negative polarity ($B_r < 0$) \citep{davis21}. For \textit{x}-mode emission, the reverse is true. The predominantly negative flux density of WX~UMa presented by \citet{davis21} therefore implies that the emission either originates from the Northern hemisphere via \textit{o}-mode emission, or from the Southern hemisphere via \textit{x}-mode emission (refer to the respective magnetic polarities in Figure~\ref{fig:zdi map}).

Provided that Equations~\ref{eq:beta condition} and \ref{eq:plasma frequency} are satisfied, an observer at a distance $d$ from the star receives a planet-induced flux density of
\begin{equation}
F_\textrm{p} = \frac{\varepsilon P}{\Omega d^2 \Delta\nu} ,
\label{eq:planet flux density}
\end{equation}
where the sign of the received flux depends on the magnetic polarity of the emitting point (the sign of ${B_x}'$), and the magnetoionic mode of the emission. Here, $\Delta\nu$ is the bandwidth of the emission, which we take as the difference between the maximum and minimum frequency along the field line where Equation~\ref{eq:plasma frequency} is satisfied. For a given field line, the bandwidth of second harmonic emission is twice that of fundamental emission. We assume that the flux density is constant with the emitted frequency, which is consistent with the observed spectra of ECMI emission on Jupiter \citep{zarka04b}.

We consider emission at the middle of the observing band here ($\nu = 144$~MHz), where $\nu$ is either $\nu_\textrm{c}$ (fundamental emission) or $2\nu_\textrm{c}$ (second harmonic emission). In Equation \ref{eq:planet flux density}, $\varepsilon$ is the fraction of the Alfv\'en wave energy that is converted into radio emission, which from observations of the Jupiter-Io sub-Alfv\'enic interaction imply that $\varepsilon = 0.01$ \citep{turnpenney18}. We use the same value in our calculations. $\Omega$ is the solid angle of the emission cone, which we compute as
\begin{equation}
\Omega = \int_0^{2\pi}\int_{\alpha - \Delta \alpha / 2}^{\alpha + \Delta \alpha / 2} \sin \theta d\theta d\phi = 4 \pi \sin (\alpha) \sin (\Delta \alpha / 2).
\end{equation}
The emission cone produced in the Jupiter-Io sub-Alfv\'enic interaction is observed to have an opening angle of $\alpha=75{\degr}$ and thickness of $\Delta\alpha=15{\degr}$ \citep{zarka04a}, giving a solid angle of $\Omega = 1.58$~sr. Again, we adopt the same value here. 

The power of the Alfv\'en waves generated by the sub-Alfv\'enic interaction is (see Appendix~\ref{sec:planet power derivation}):
\begin{equation}
P = \pi^{1/2} {R_\textrm{m}}^2 B_\textrm{w} \rho_\textrm{w}^{1/2} \Delta u^2 \sin^2\theta .
\label{eq:planet power}
\end{equation}
Here, $R_\textrm{m}$ is the effective radius of the planet, which in the case of a magnetised planet, can be approximated as the size of its magnetopause. The magnetopause extends out to where the magnetic pressure of the planet equals that of the incident stellar wind, which is predominantly magnetic in the sub-Alfv\'enic region. Therefore, with an incident stellar wind with a magnetic field strength of $B_\textrm{w}$, and a planetary field strength at the magnetopause of $B_\textrm{p,m}$, pressure balance gives \citep{vidotto13}:
\begin{equation}
\frac{{B_\textrm{w}}^2}{8\pi} \simeq \frac{{B_\textrm{p,m}}^2}{8\pi} .
\label{eq:planet pressure balance}
\end{equation}
For a dipolar magnetic field that is aligned with the stellar rotation axis, the strength of the planetary field at the magnetopause is
\begin{equation}
B_\textrm{p,m} = \frac{B_\textrm{p}}{2} \Big( \frac{R_\textrm{p}}{R_\textrm{m}} \Big)^3 ,
\end{equation}
where $B_\textrm{p}$ and $R_\textrm{p}$ are the planetary polar magnetic field strength and radius. Plugging this in to Equation~\ref{eq:planet pressure balance}, we can obtain an expression for the magnetopause size:
\begin{equation}
R_\textrm{m} = \Big( \frac{1}{2} \frac{B_\textrm{p}}{B_\textrm{w}} \Big)^{1/3} R_\textrm{p} .
\label{eq:planet magnetopause size}
\end{equation}
As the large-scale magnetic field of WX~UMa resembles an aligned dipole, the equatorial field strength exhibits small variations in a circular orbit around the star. Combined with the weak dependence on the stellar wind magnetic field strength in Equation~\ref{eq:planet magnetopause size}, the size of the magnetopause is effectively constant for a given orbital distance.

In terms of the remaining terms in Equation~\ref{eq:planet power}, $\rho_\textrm{w}$ is the density of the stellar wind at the position of the planet respectively, and $\Delta u$ is the relative velocity between the stellar wind and planet. At a distance $a$ from the star, we assume that the planet is in a circular orbit in the equatorial plane of the star, moving in the positive azimuthal direction $\boldsymbol{\hat{\phi}}$ (prograde orbit). The planet orbits with a Keplerian velocity of $\boldsymbol{u}_\textrm{p} = \sqrt{G M_\star / a}\boldsymbol{\hat{\phi}}$, where $G$ is the gravitational constant. The relative velocity between the stellar wind and planet is given by $\boldsymbol{\Delta u} = \boldsymbol{u}_\textrm{w} - \boldsymbol{u}_\textrm{p}$, and $\theta$ is the angle between the vectors $\boldsymbol{\Delta u}$ and $\boldsymbol{B}_\textrm{w}$. 

\begin{figure}
\includegraphics[width = \columnwidth]{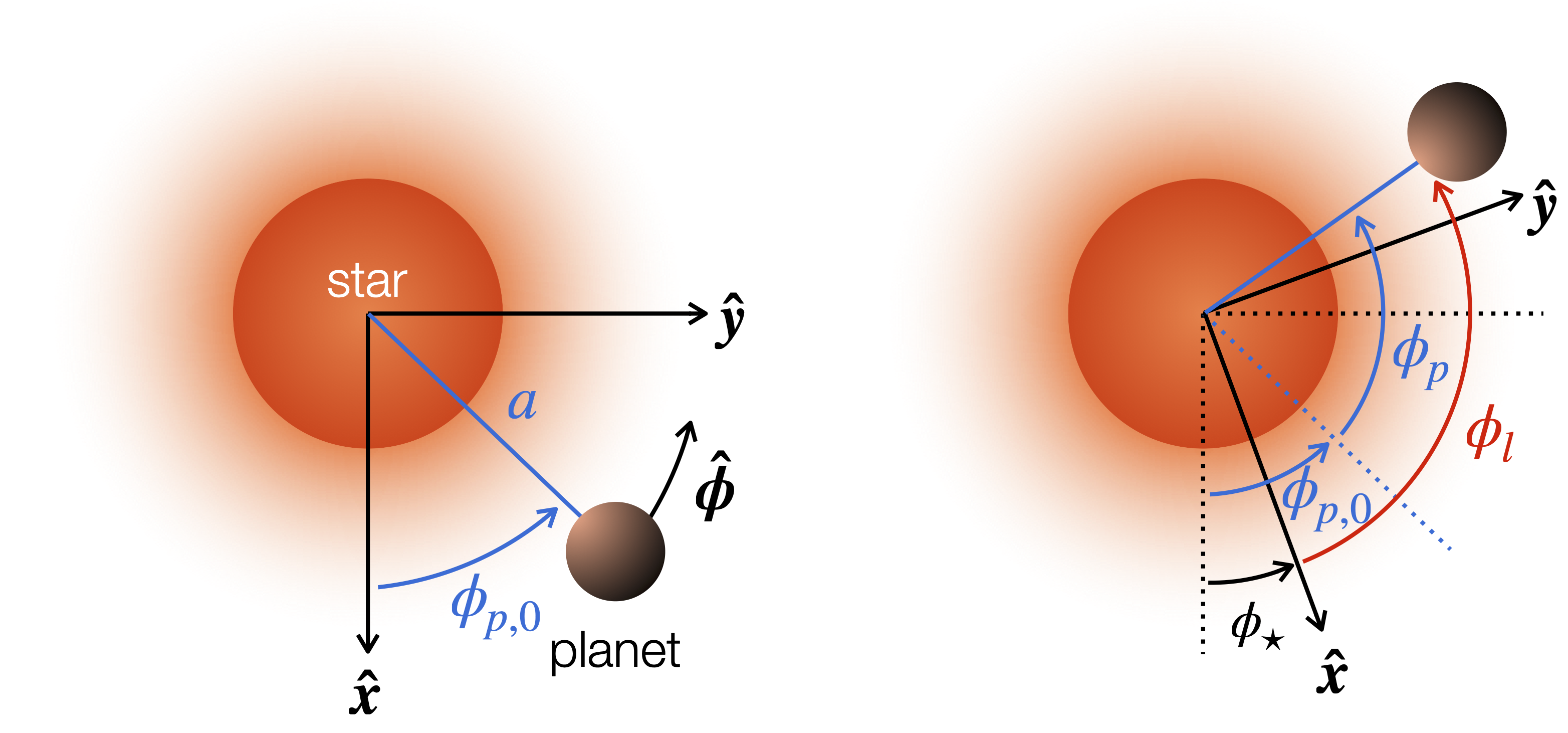}
\caption{Top-down view of the equatorial plane of the star, defined by vectors $\boldsymbol{\hat{x}}$ and $\boldsymbol{\hat{y}}$. The rotation axis points out of the figure. The planet orbits in the direction of the vector $\boldsymbol{\hat{\phi}}$. Initially, it is at an orbital phase of $\phi_{\textrm{p},0}$, at a distance $a$ from the star (left). Then, the star rotates by phase $\phi_\star$, and the planet progresses by phase $\phi_\textrm{p}$ (right). The phase $\phi_l$ is the longitude of the magnetic field line in the stellar coordinate system that the planet intercepts as it orbits the star. For clarity, we do not show the planet's own magnetosphere.}
\label{fig:sketch planet orbit geometry}
\end{figure}

We assume that the planet has an initial orbital phase of $\phi_{\textrm{p},0}$ at the start of the stellar rotation phase covering the beginning of the 2014 radio observations. The star subsequently progresses by rotation phase $\phi_\star$, and the planet phase increases by $\phi_\textrm{p}$. A sketch of the geometry described here is shown in Figure~\ref{fig:sketch planet orbit geometry}. At the stellar rotation phase $\phi_\star$, the planet intercepts the magnetic field line at longitude
\begin{equation}
\phi_l = \phi_{\textrm{p},0} + \phi_\textrm{p} - \phi_\star = \phi_{\textrm{p},0} + \Big( \frac{P_\star}{P_\textrm{p}} - 1 \Big) \phi_\star .
\label{eq:planet fieldline longitude}
\end{equation}
In the stellar coordinate system, Equation~\ref{eq:planet fieldline longitude} describes the position of the planet in the equatorial plane of stellar wind as a function of stellar rotation. The two phases $\phi_\star$ and $\phi_\textrm{p}$ relate to one another via the stellar rotation and orbital periods $P_\star$ and $P_\textrm{p}$:
\begin{equation}
\phi_\textrm{p} = \frac{P_\star}{P_\textrm{p}} \phi_\star ,
\end{equation}
and $P_\textrm{p}$ is given by Kepler's third law:
\begin{equation}
P_\textrm{p} \approx 2\pi\sqrt{\frac{a^3}{GM_\star}} .
\end{equation}

\subsection{Can the emission from WX UMa be explained by an undetected planet?}
\label{sec:best fit}

Our goal here is to determine whether the observed emission from WX UMa presented by \citet{davis21} can be reproduced by a planet orbiting in the sub-Alfv\'enic regime of the stellar wind. If so, what are the planetary and orbital parameters that best-reproduce the radio emission of the star? In our model, the planet's magnetopause size $R_\textrm{p}$, orbital distance $a$, and initial orbital phase $\phi_{\textrm{p},0}$ are free parameters. For orbital distances $\lesssim5~R_\star$, fundamental ECMI emission cannot be generated, as the magnetic field strength of each line exceeds that required for fundamental ECMI at 144~MHz (the middle of the observing band). Similarly, second harmonic ECMI emission is limited to orbital distances $\gtrsim7~R_\star$. At $\sim80~R_\star$, the stellar wind also becomes super-Alfv\'enic, and for orbits greater than this distance, an orbiting planet can no longer induce radio emission from the star. Therefore, we limit our range of orbital distances to 5--80~$R_\star$ for fundamental ECMI, and 7--80~$R_\star$ for second harmonic ECMI respectively.

Within these ranges of orbital distances, we vary the value of $\phi_{\textrm{p},0}$ from 0 to 1, and find the point on the magnetic field line connecting to this location in the orbital plane that corresponds to fundamental and second harmonic ECMI emission via Equation~\ref{eq:cyclotron frequency}. For each position of the planet, we linearly interpolate the relevant stellar wind properties to compute the power of the Alfv\'en waves generated via Equation~\ref{eq:planet power}, as well as those needed to determine the beaming angle at a given frequency on the connecting field line via Equation~\ref{eq:beta}. Provided that the emission from each point is beamed towards the observer (Equation~\ref{eq:beta condition}), and the cyclotron frequency exceeds the local plasma frequency (Equation~\ref{eq:plasma frequency}), we compute the flux density that the observer sees for a range of radii using Equation~\ref{eq:planet flux density}. We also consider emission in both the \textit{x} and \textit{o}-mode, accounting for the respective sign of the flux density, which depends on the orientation of the magnetic field relative to the observer (see Section~\ref{sec:planet radio emission}). We perform our calculations for all stellar rotation phases covered by the radio observations of WX UMa, using Equation~\ref{eq:planet fieldline longitude} to determine the magnetic field line the planet intercepts as the star rotates. This gives us the flux density of fundamental and second harmonic ECMI emission induced by the planet as a function of stellar rotation, which we refer to as the `radio lightcurve' of the star.

To determine which set of parameters best-reproduce the observed radio lighcurve of WX~UMa, we use a $\chi^2$ test as a guide. We calculate the value of $\chi^2$ as
\begin{equation}
\chi^2 = \sum \frac{(F_\textrm{obs} - F_\textrm{p})^2}{n_\textrm{obs}{\sigma_\textrm{obs}}^2} ,
\label{eq:chi2}
\end{equation}
where $F_\textrm{obs}$ and $\sigma_\textrm{obs}$ are the observed flux and its respective error, $n_\textrm{obs}$ is the number of observed flux values, and $F_\textrm{p}$ is the planet-induced flux density computed using Equation~\ref{eq:planet flux density}. The lightcurve presented in \citet{davis21} is binned in to 2-hour windows, which is too coarse for comparison to our model. Therefore, following the same procedure described in \citet{davis21}, we extract the 144~MHz lightcurve with a temporal resolution of $\sim20$ minutes, giving a total number of observations of $n_\textrm{obs} = 72$ (24 for each epoch).

We compute the value of $\chi^2$ for each set of inputs, and then visually check the lightcurves that produce values of $\chi^2\approx 1$. For fundamental ECMI emission, we find that the observed lightcurve of WX~UMa is best-reproduced by a planet orbiting the star at $\sim0.034$~au ($60.6~R_\star$), with a $\chi^2$ value of 0.97. The magnetoionic mode we identify here is the \textit{o}-mode, with the induced emission occurring in the Northern hemisphere. The initial orbital phase of the planet relative to the 2014 radio observations is 0.15, and the corresponding magnetopause size for these values is $11.9~R_\textrm{Nep}$. For reference, the size of Jupiter's magnetopause varies from $\sim50$ to 100 Jupiter radii depending on the solar wind conditions \citep{bagenal13}. Using Equation~\ref{eq:planet magnetopause size}, the resulting planetary radius is 0.9 -- 1.9~Neptune radii ($R_\textrm{Nep}$) for polar magnetic field strengths from 10 –- 100~G, which is the range of inferred field strengths for hot Jupiters \citep[Table 3 of][neglecting their values for an efficiency of 100\%]{cauley19}.

For second harmonic ECMI emission, the planet that best-reproduces the observed emission also orbits at $\sim0.034$~au, with an initial orbital phase of 0.13 and a corresponding $\chi^2$ of 0.92. Again, we identify the emission to occur in the \textit{o}-mode here. The size of the magnetopause is $16.6~R_\star$, which is larger than that for fundamental ECMI, to compensate for the larger bandwidth $\Delta \nu$ in Equation~\ref{eq:planet flux density} (for a given field line, the bandwidth of second harmonic emission is twice that of fundamental emission). For polar magnetic field strengths of 10 -- 100~G, the planetary radius is 1.3 -- 2.7~$R_\textrm{Nep}$. We list the orbital phases of the planet in each case across the 3 observing epochs in Table~\ref{tab:planet phases}.

\begin{figure*}
\includegraphics[width = \textwidth]{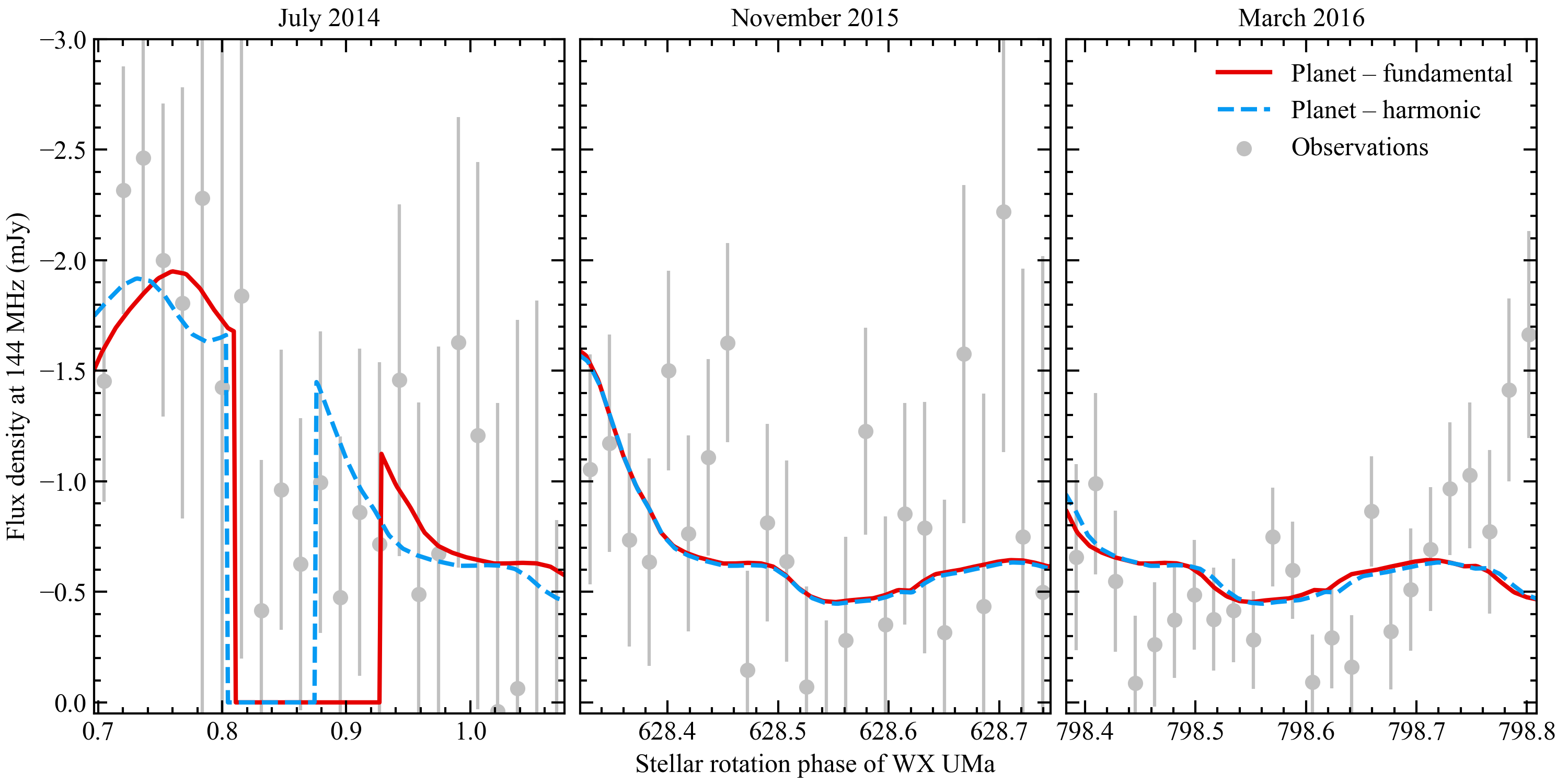}
\caption{Best-fitting planet-induced radio lightcurves of WX~UMa at 144~MHz, for fundamental (red) and second harmonic (blue) ECMI emission. In both cases, the predicted planet orbits at $\sim0.034$~au, with a magnetopause size of 11.9 and 16.6~$R_\textrm{Nep}$ respectively. For planetary magnetic field strengths of 10 -- 100 G, the radius of the planet ranges from 0.9 to 1.9~$R_\textrm{Nep}$ in the case of fundamental ECMI, and 1.3 to 2.7~$R_\textrm{Nep}$ for second harmonic ECMI. The grey dots and lines show the observed flux density and respective $1\sigma$ errors from WX~UMa at a temporal resolution of $\sim20$ minutes.}
\label{fig:lightcurves}
\end{figure*}

In Figure~\ref{fig:lightcurves}, we show a comparison of the best-fitting planet-induced lightcurves for fundamental and second harmonic ECMI to the observed emission from WX~UMa presented in \citet{davis21}. The planet-induced flux densities obtained are predominantly negative, in agreement with the observations and the assumption that the emission occurs in the \textit{o}-mode. The surface field of the star exhibits a predominantly negative polarity in the North pole. As a result, the open field lines connecting to the planetary orbits identified tend to connect back to the North pole as opposed to the South pole, producing a negative flux density.

We note that there is not a unique set of best-fitting values; rather, those presented here represent those that produce the lowest value of $\chi^2$. We find a few additional values that produce reasonable fits to the observations. These typically have values of $\chi^2 \lesssim 1.3$. We list all sets of best-fitting values for fundamental and second harmonic emission in Table~\ref{tab:best fits}.

\begin{table}
\caption{Orbital phases of the predicted planet that best-reproduces the observed lightcurve of WX~UMa at 144~MHz during the 3 observing epochs, for fundamental and second harmonic ECMI. The values shown here are $\phi_{\textrm{p},0} + \phi_\textrm{p}$ (see Figure~\ref{fig:sketch planet orbit geometry}).}
\label{tab:planet phases}
\centering
\begin{tabular}{lccc}
\hline
& July 2014 & November 2015 & March 2016 \\
\hline
Fundamental & 0.04 -- 0.08 & 0.49 -- 0.54 & 0.50 -- 0.54 \\
Harmonic & 0.88 -- 0.92 & 0.36 -- 0.40 & 0.37 -- 0.42 \\
\hline
\end{tabular}
\end{table}

\begin{table}
\caption{Predicted planetary and orbital parameters that reproduce the observed lightcurve of WX~UMa for fundamental and second harmonic ECMI emission, listed in order of increasing orbital distance. The values from left to right are the planetary orbital distance ($a$), orbital period ($P_\textrm{p}$), magnetopause size ($R_\textrm{m}$), radius assuming a polar magnetic field strength of 10~G ($R_\textrm{p,10 G}$) and 100~G ($R_\textrm{p,100 G}$) calculated using Equation~\ref{eq:planet magnetopause size}, initial orbital phase $\phi_{\textrm{p},0}$, and resulting $\chi^2$ calculated using Equation~\ref{eq:chi2}. Values of $\chi^2\lesssim 1.3$ generally produce reasonable results.}
\label{tab:best fits}
\centering
\begin{tabular}{ccccccc}
\hline
$a$ & $P_\textrm{p}$ & $R_\textrm{m}$ & $R_\textrm{p,10 G}$ & $R_\textrm{p,100 G}$ & $\phi_{\textrm{p},0}$ & $\chi^2$ \\
(au) & (days) & ($R_\textrm{Nep}$) & ($R_\textrm{Nep}$) & ($R_\textrm{Nep}$) & & \\
\hline
\multicolumn{5}{l}{\underline{Fundamental}:} \\
0.028 & 5.4 & 8.4 & 1.6 & 0.7 & 0.13 & 1.20 \\
0.033 & 7.2 & 9.3 & 1.5 & 0.7 & 0.15 & 1.32 \\
0.034 & 7.4 & 11.9 & 1.9 & 0.9 & 0.15 & 0.97 \\
\multicolumn{5}{l}{\underline{Harmonic}:} \\
0.028 & 5.4 & 14.7 & 2.7 & 1.3 & 0.10 & 1.08 \\
0.034 & 7.4 & 16.6 & 2.7 & 1.3 & 0.13 & 0.92 \\
\hline
\end{tabular}
\end{table}


\subsection{Visibility of the potential planet}
\label{sec:planet visibility}

With the orbital and planetary parameters identified for a planet that could reproduce the observed emission from WX~UMa, the next question is how visible would this planet be at different wavelengths? Here we explore the visibility of the planetary signatures, both in the radio and using traditional exoplanet detection methods.


\subsubsection{Radio visibility at 144~MHz}

First, in the radio we compute the planet-induced lightcurve at 144~MHz using the best-fitting values for fundamental and second harmonic ECMI emission identified in Section~\ref{sec:best fit}, covering the planetary orbit over 1000 stellar rotations. In both cases, we find that if we observe the system at randomly chosen 8-hour window, we catch the planetary signal 26\% of the time when accounting for the emission beaming and plasma frequency condition (Equations~\ref{eq:beta condition} and \ref{eq:plasma frequency}). The average flux density induced via fundamental ECMI that is visible to the observer is $-0.85$~mJy, with a $1\sigma$ standard deviation of 0.50~mJy. For second harmonic ECMI, the average flux visible to the observer is $-0.85\pm0.54$~mJy. For reference, 8~hour observations with LOFAR typically reach a sensitivity level of $\sim0.1$~mJy. Considering that the 144~MHz emission from WX~UMa is seen nearly continuously over the three separate epochs may make it difficult to reconcile with the planet-induced case presented here. That being said, a much longer radio observing campaign will be needed to assess this further. Additionally, propagation effects such as free-free/gyroresonance absorption and refraction may also further hinder the detection/visibility of a planet-induced signal (see Appendix~\ref{sec:propagation quantities}).

\begin{figure}
\includegraphics[width = \columnwidth]{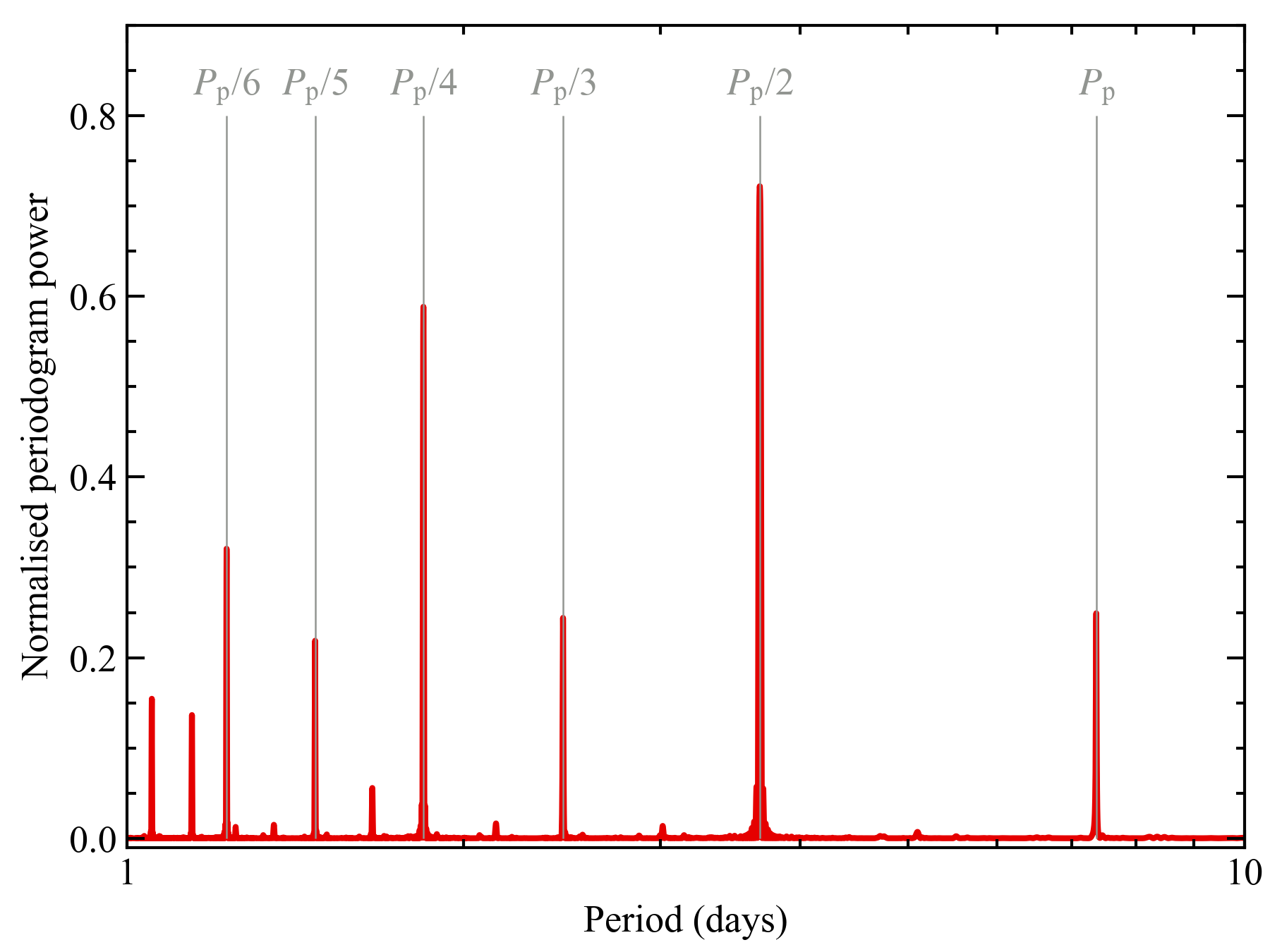}
\caption{Periodogram of the planet-induced radio lightcurve at 144~MHz from WX~UMa via fundamental ECMI. The lightcurve shows prominent periodicity at the fundamental and harmonics of the planetary orbital frequency $\nu_\textrm{p} = 1 / P_\textrm{p}$, with the most dominant peak being found at the second harmonic. These are indicated by the vertical lines.}
\label{fig:periodogram}
\end{figure}

We then investigate what the dominant periods are in the planet-induced radio signal, which is useful information for planning future radio observations of the star. To explore this, we employ the Lomb-Scargle periodogram \citep{lomb76, scargle82}. In Figure~\ref{fig:periodogram}, we show the periodogram of the planet-induced radio signal at 144~MHz for fundamental ECMI emission. The periodogram of the lightcurve induced via second harmonic ECMI gives the same result, as the orbital distances in both cases are equivalent. We find prominent peaks at the fundamental and harmonics of the orbital frequency of the planet $\nu_\textrm{p} = 1/P_\textrm{p}$, with the most dominant peak occurring at the second harmonic ($2\nu_\textrm{p}$). This can be interpreted as when the planet intercepts field lines that are in the plane of the sky, which occurs twice per orbit. Such information could prove useful in carrying out future searches for such signals in the radio. In contrast, chromospheric emission lines such as Ca II H \& K (3968.47 \& 3933.66~\AA) and He I D3 (5875.62~\AA), which are thought to be tracers of star-planet interactions, are expected to show strong periodicity with the fundamental and harmonics of the beat period of the star and planet ($P_\textrm{beat} = P_\star P_\textrm{p} / |P_\star - P_\textrm{p}|$) and the fundamental of the planetary orbital period \citep{fischer19, klein22}. This highlights the benefits of simultaneous multi-wavelength observations (e.g. optical and radio) in probing star-planet interactions.

\subsubsection{Planet-induced dynamic radio spectrum of WX UMa}

\begin{figure*}
\includegraphics[width = \textwidth]{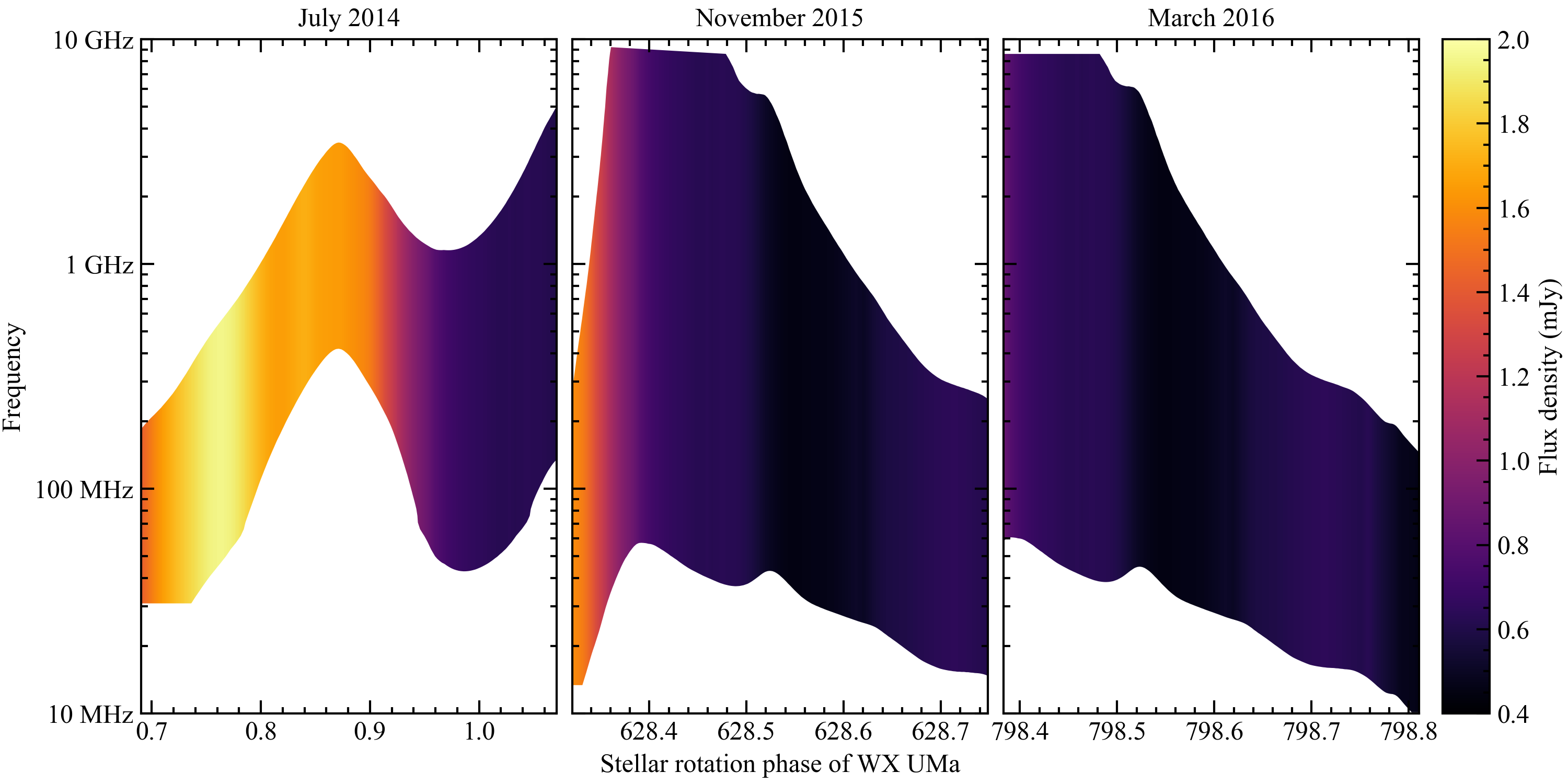}
\caption{Dynamic radio spectra of WX~UMa induced by the planet that best-reproduces the observed emission via fundamental ECMI at 144~MHz over the three epochs (left to right). At 144~MHz, the morphology of the flux density is that shown in Figure~\ref{fig:lightcurves} for fundamental ECMI. The radio emission is only visible to the observer if it is beamed along the line of sight, and if the local cyclotron frequency exceeds the plasma frequency. Note here that we have neglected the sign of the flux density. In the LOFAR convention, this emission would be seen as a negative flux (from the Northern hemisphere).}
\label{fig:dynamic spectrum - fundamental}
\end{figure*}

We now explore the range of possible emission frequencies along the entire magnetic field line connecting to the planet at each point in its orbit during the LOFAR observations. Provided Equation~\ref{eq:plasma frequency} is satisfied, and the emission is beamed towards the observer, we compute the flux density emitted from each point on the field line from 10~MHz to 10~GHz for fundamental ECMI emission, producing a dynamic radio spectrum at each of the three epochs. This frequency range covers all possible emitting frequencies via fundamental ECMI for the stellar magnetic field lines connecting to the predicted planetary orbit at $\sim0.034$~au, which is determined by the field strength along each line (see Equation~\ref{eq:cyclotron frequency}). The resulting dynamic spectrum is shown in in Figure~\ref{fig:dynamic spectrum - fundamental}. As the best-fitting orbital distance for second harmonic ECMI emission is the same as that for fundamental emission, the resulting dynamic spectrum is very similar, albeit over a bandwidth that is twice that of fundamental emission. This is shown in Figure~\ref{fig:dynamic spectrum - harmonic}. While both are comparable, detection of second harmonic emission may be more favourable than fundamental emission, as stellar winds are more optically thin at higher wavelengths in the radio regime \citep[][see Appendix~\ref{sec:propagation quantities} also]{panagia75, ofionnagain19, kavanagh19}. Multi-wavelength radio observations of the system could also help distinguish between the two modes \citep[e.g.][]{das21}.

\subsubsection{Transit of the stellar disk}

In terms of traditional exoplanet detection methods, for a planet to transit WX~UMa, its orbital distance $a$ must satisfy
\begin{equation}
a < \frac{R_\star + R_\textrm{p}}{\sin i_\star} .
\end{equation}
With a stellar inclination of $i_\star = 40\degr$ \citep{morin10} and a maximum radius of $\sim2.7~R_\textrm{Nep}$, the planet would need to orbit WX~UMa at a distance of $\lesssim2.8~R_\star$ (0.002~au), assuming it orbits in the equatorial plane. This value is smaller than the minimum orbital distance where fundamental or second harmonic ECMI emission can be induced in the star. Therefore, with our predicted orbital parameters, such a planet would not be detectable via transit observations.


\subsubsection{Radial velocity signatures}

While detectability via transits is not likely for the putative planet, it may be possible for it to generate measurable radial velocity signatures in spectroscopic lines from the star. The radial velocity semi-amplitude due to the presence of a planet is \citep{wright09}:
\begin{equation}
K = \Big(\frac{G}{a}\Big)^{1/2} \frac{M_\textrm{p}\cos i_\star}{(M_\star + M_\textrm{p})^{1/2}} ,
\end{equation}
where $M_\textrm{p}$ is the mass of the planet. Assuming a Neptunian density of 1.6~g~cm$^{-3}$, for a planetary radius of 0.7 -- 2.7~$R_\textrm{Nep}$ the semi-amplitude varies from 7 -- 396~m~s$^{-1}$. A signal of this strength is well within the detection capabilities of current-generation spectrographs. However, radial velocity measurements of the star are likely to be completely dominated by stellar activity, with the associated radial velocity jitter ranging from 60 -- 530~m~s$^{-1}$ \citep{morin10}.


\section{Reconnection at the edge of the magnetosphere}

ECMI emission may also be generated from WX~UMa without the presence of a planet. If reconnection occurs in the current sheet of the stellar wind plasma at the edge of the magnetosphere \citep{linsky92, trigilio04, nichols12}, electrons may be accelerated towards the magnetic polar regions. This acceleration could produce a population of high-velocity electrons in a loss-cone or horseshoe distribution, which in turn can power ECMI (see Section~\ref{sec:intro}). Provided there is no evolution of the magnetic field or surrounding stellar wind plasma, emission generated via such a mechanism would be modulated by the stellar rotation period.

To investigate this scenario for WX~UMa, we first identify the magnetic field lines connecting to the Alfv\'en surface in the middle of the current sheet (where the radial component of the magnetic field is zero). We then find the points along each field line which correspond to fundamental and second harmonic ECMI emission at 144~MHz using Equation~\ref{eq:cyclotron frequency}. At each point, we assume that the emitted flux is proportional to an emission proxy $p$, which we normalise by its maximum value at the emitting points. We calculate the total flux as the sum the contributions from all emitting points at 144~MHz:
\begin{equation}
F = C \sum p ,
\label{eq:flux currents}
\end{equation}
where $C$ is a scaling constant. Again, we check the same beaming and plasma frequency conditions described by Equations~\ref{eq:beta condition} and \ref{eq:plasma frequency}.

We consider two proxies for the emission in the closed field of WX~UMa. The first follows the formulations of \citet{llama18}, who illustrated that the observed radio modulation at 4 -- 8~GHz of the active M~dwarf V374~Peg can be reproduced by assuming the emission is proportional to the density of free electrons in the closed field. The second proxy utilised is the current density along the field lines. This is inspired by Ohmic heating, in which the energy is proportional to the square of the current density \citep{meyer13, gibb14}. Ohmic heating is thought to originate in reconnection events \citep{benz17}, such as those which may power ECMI from magnetised stars.

Utilising these two proxies, we carry out the same process as for the planet-induced scenario. By varying the scaling constant $C$ in Equation~\ref{eq:flux currents}, we determine what values minimises value of $\chi^2$:
\begin{equation}
\chi^2 = \sum \frac{(F_\textrm{obs} - F)^2}{n_\textrm{obs}{\sigma_\textrm{obs}}^2} .
\end{equation}
As we assume that the large-scale magnetic field and plasma density do not change over the course of the radio observations, we phase-fold the data in our analysis here. Considering both proxies, combined with emission generated in both the \textit{x} and \textit{o}-mode via fundamental and second harmonic ECMI at 144~MHz, we find that only the density proxy can reproduce the observed emission via fundamental \textit{o}-mode ECMI. The maximum density that we normalise the proxy by for the points emitting fundamental ECMI at 144~MHz is $1.56\times10^9$~cm$^{-3}$. The scaling constant obtained for the density proxy which best-reproduces the observed emission is $C = 0.21$~mJy, with a corresponding $\chi^2$ of 2.03.

\begin{figure}
\includegraphics[width = \columnwidth]{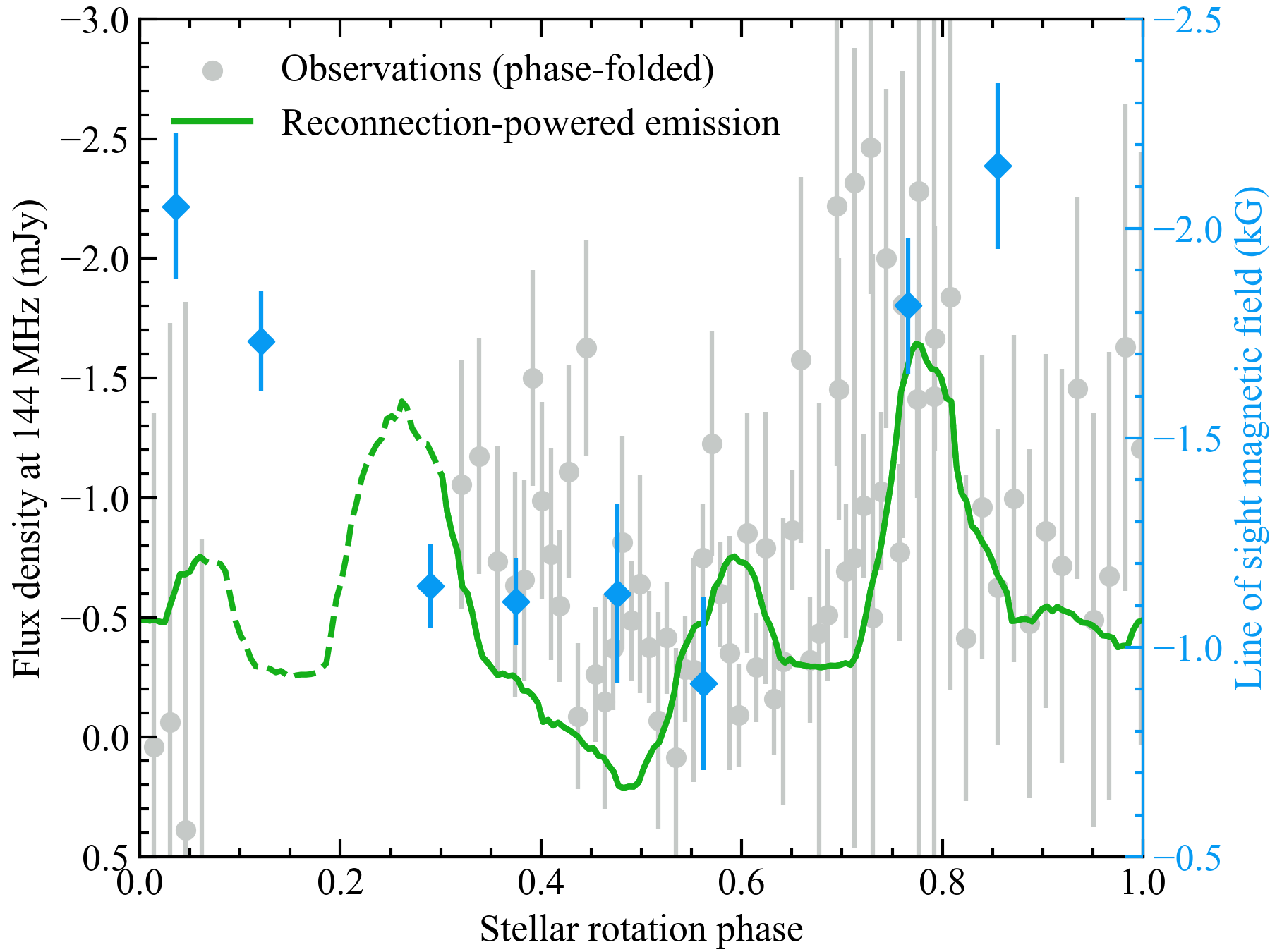}
\caption{Comparison of the phase-folded radio observations of WX~UMa at 144~MHz (grey dots) to that produced from the magnetosphere using the density of free electrons as the emission proxy (green line). The emission here is assumed to be generated in the \textit{o}-mode via fundamental ECMI. The vertical bars show the $1\sigma$ errors associated with the observed flux densities. We do not consider the stellar rotation phases for the dashed part of the green line in our analysis, as there is no data available in this region. We also overlay the line of sight magnetic field strength from \citet{morin10} as blue diamonds. These values seem to follow the modulation of the phase-folded radio lightcurve relatively well between rotation phases 0.3 -- 0.7.}
\label{fig:reconnection}
\end{figure}

In Figure~\ref{fig:reconnection} we show a comparison of the phase-folded radio observations to that obtained using the density as a proxy for the emission. As the emission depends only on the structure of the magnetosphere, it is modulated by the stellar rotation period. This distinguishes it from planet-induced emission, which has prominent periodicity with harmonics of the planetary orbital frequency (e.g.~Figure~\ref{fig:periodogram}). We also overlay the line of sight magnetic field strength from Table~A4 in \citet{morin10} in Figure~\ref{fig:reconnection} to compare to the phase-folded lightcurve. The modulation of these values seem to match the observed emission relatively well for rotation phases 0.3 -- 0.7.

Visually, using the density as a proxy for the emission reproduces the observed phase-folded lightcurve relatively well. That being said, the drawback of the model here is that the value obtained for the scaling constant is not physically motivated, unlike those identified in the planet-induced scenario, which describe a sub-au orbit of a Neptune-sized planet around the star. Additionally, the fit to the data is poorer here than in the planet-induced scenario. However, we expect the planet-induced signal to be modulated primarily by harmonics of the orbital period of the potential planet (7.4~days), whereas for the reconnection scenario, the emission would be modulated with the stellar rotation period (0.78~days). A more stringent comparison of the two models therefore requires radio observations of the star covering multiple planetary orbits. Nevertheless, these proxies illustrate how the magnetospheric structure of the star may imprint itself in the resulting radio emission. Again, in future adaptations of a model such as this, accounting for the absorption and refraction of the generated emission may be necessary (see Appendix~\ref{sec:propagation quantities}). We also note that there have recently been numerous arguments made in the literature disfavouring a mechanism such as this in powering auroral emission on Jupiter, which is the inspiration for this scenario \citep{bonfond20}. It is also worth mentioning that recent work has illustrated that incoherent non-thermal radio emission from Jupiter's radiation belt is in agreement with that expected to be generated in the magnetospheres of hot B/A-type stars \citep{leto21, owocki22}.

\section{Discussion \& Conclusions}

\subsection{Model limitations}
\label{sec:model limitations}

One caveat to the application of our models to WX~UMa is the time interval between the observations used to reconstruct the stellar surface magnetic field and the radio observations. Our stellar wind simulation presented in Section~\ref{sec:wind model} is effectively a snapshot of the wind of the star in April~2006, which is the epoch of the spectropolarimetric observations used by \citet{morin10} to reconstruct the surface field map implemented in our stellar wind model. The first LOFAR observations of the star then occurred in July~2014. Within this time, the magnetic field of the star could have evolved significantly. \citet{morin10} also derived a rotation period of $0.78\pm0.20$~days for star from the 2006 observations. Fast-forwarding to the start of the 2014 radio observations, the star had undergone over 3,800 rotations. Propagating the error in rotation period, the error in the stellar rotation phase in July~2014 is $\pm99$. As a result, there is a large uncertainty in the true rotation phase of the star during the radio observations. In order to mitigate this effect, near-simultaneous spectropolarimetric and radio observations and higher precision stellar rotation periods are needed.

Another set of aspects which are not addressed in our model are propagation effects on the escaping radio emission. These are not limited to refraction \citep{melrose82, trigilio11, das20}, dispersion and scattering \citep{gudel91}, reflections off of density boundary layers in the corona \citep{melrose06}, and gyroresonance absorption, which is thought to be significant in the coronae of M~dwarfs \citep{stepanov01, vedantham21}. Low-frequency radio emission can also be readily absorbed via free-free absorption \citep{rybicki86, kavanagh20}. In Appendix~\ref{sec:propagation quantities}, we show both the refractive indices of \textit{x} and \textit{o}-mode emission, as well as the optical depth at 144~MHz. While self-consistently accounting for these processes is currently beyond the scope of this paper, their inclusion is worth exploring in the future.

There are two other parameters in our model which we did not explore in this work: the properties of the emission cone, and the conversion efficiency of the Alfv\'en waves produced by the star-planet sub-Alfv\'enic interaction to radio emission. These two parameters depend on the velocity of the accelerated electrons, which in turn is dependent on the conversion of the Alfv\'en wave energy. As this process of Alfv\'en wave generation via the sub-Alfv\'enic star-planet interaction is not explicitly calculated in our stellar wind simulation, we cannot determine appropriate values for these parameters in a self-consistent way.


\subsection{Conclusions}

In this paper, we presented a refined model for planet-induced radio emission from low-mass stars, based on its initial form developed by \citet{kavanagh21}. This model utilises the realistic information about the large-scale stellar magnetic field and plasma environment that is provided by stellar wind simulations, and accounts for both the beaming and polarisation of the radio emission, as well as the respective stellar rotational and planetary orbital motion. To the best of our knowledge, it is the first of its kind in the literature.

Applying our updated model to the M~dwarf WX~UMa, we have shown that a Neptune-sized exoplanet with a magnetic field strength of 10 -- 100~G orbiting the star at $\sim0.034$~au can accurately reproduce recent radio observations of the star at 144~MHz. If the recent observations are indeed of a star-planet interaction origin, the signal with a strength of up to 2~mJy may be visible up to 26\% of the time at 144~MHz, with the emission ranging from 10~MHz up to 20~GHz, depending on if the emission is generated at the fundamental or second harmonic of the local cyclotron frequency. While transits of our putative planet are unlikely due to the stellar inclination, the induced semi-amplitudes well within the detection capabilities of current-generation spectrographs are possible. That being said, the activity of the star may hinder such detections.

An interesting question arising from this work is how to distinguish between a planet inducing fundamental vs.~second harmonic ECMI emission, as both can produce highly circularly polarised emission \citep{vedantham21}. Across the range of 10~MHz--20~GHz, the morphology of the emission does not differ significantly, however second harmonic emission is more prominent at higher frequencies by nature. If a planet-induced radio signature was identified, multi-wavelength radio observations would certainly aid disentangling the emission harmonic. In addition to this, fundamental ECMI emission may be more susceptible to being attenuated significantly as it propagates out of the system. As a result, detection may be more favourable in the case of second harmonic emission. \citet{melrose82} also illustrated that for a relative cool ($<10^7$~K) and rarefied plasma ($<3\times10^8$~cm$^{-3}$), emission at the second harmonic can escape more easily than fundamental emission, the latter of which being subjected to significant gyroresonance absorption (see Appendix~\ref{sec:propagation quantities}).

We also developed a model to explore an alternative proposed mechanism for generating ECMI in magnetised low-mass stars. This mechanism relies on persistent reconnection occurring at the edge of the stellar magnetosphere. We utilised the information obtained from our wind simulation about the large-scale magnetic field, and explored different emission proxies to mimic emission from the field lines associated with this region. We found that assuming the emission scales with the density of free electrons, we can reproduce the observed emission at 144~MHz to an extent, albeit with a poorer fit to the data than in the planet-induced scenario. However, we note here that there is no precedent for the associated scaling prescriptions obtained in this analysis. Future work will be needed to provide more realistically account for the energy available to power such mechanisms. 

It is worth noting the differences between emission induced by a planet vs.~that generated in the reconnection scenario. Over many orbital periods of the planet, there is a dominant periodicity at the fundamental and harmonics of the orbital frequency. This signature would easily distinguish it from the alternative proposed mechanism, which would show regular modulation with the stellar rotation period (assuming the magnetic field does not evolve significantly). Distinguishing between the two however would likely need a more extensive radio observing campaign.

To conclude, the planet-induced radio emission model presented here has fantastic potential for application to current and upcoming radio observations, to be used as a new tool to discover planets around nearby stars at radio wavelengths. However, alternative scenarios such as reconnection at the edge of the magnetosphere should also be explored further. Future application of our planet-induced model to radio observations of low-mass stars such as those presented by \citet{callingham21b}, as well as from upcoming developments such as the third phase of the Owens Valley Long Wavelength Array \citep[OVRO-LWA,][]{hallinan15} and FARSIDE lunar array \citep{hallinan21} will be of great use in guiding follow-up observations using traditional exoplanet detection techniques.


\section*{Acknowledgements}

We thank the anonymous referee for their insightful comments and suggestions. RDK acknowledges funding received from the Irish Research Council (IRC) through the Government of Ireland Postgraduate Scholarship Programme. RDK and AAV acknowledges funding from the European Research Council (ERC) under the European Union's Horizon 2020 research and innovation programme (grant agreement No 817540, ASTROFLOW). We acknowledge the provisions of the Space Weather Modelling Framework (SWMF) code from the Center for Space Environment Modeling (CSEM) at the University of Michigan, and the computational resources of the Irish Centre for High End Computing (ICHEC), both of which were utilised in this work.


\section*{Data availability}

The data presented in this paper will be shared on reasonable request to the corresponding author.


\bibliographystyle{mnras}
\bibliography{bibliography}


\appendix

\section{Stellar wind plasma properties}
\label{sec:plasma properties}

\begin{figure*}
\centering
\includegraphics[width = \textwidth]{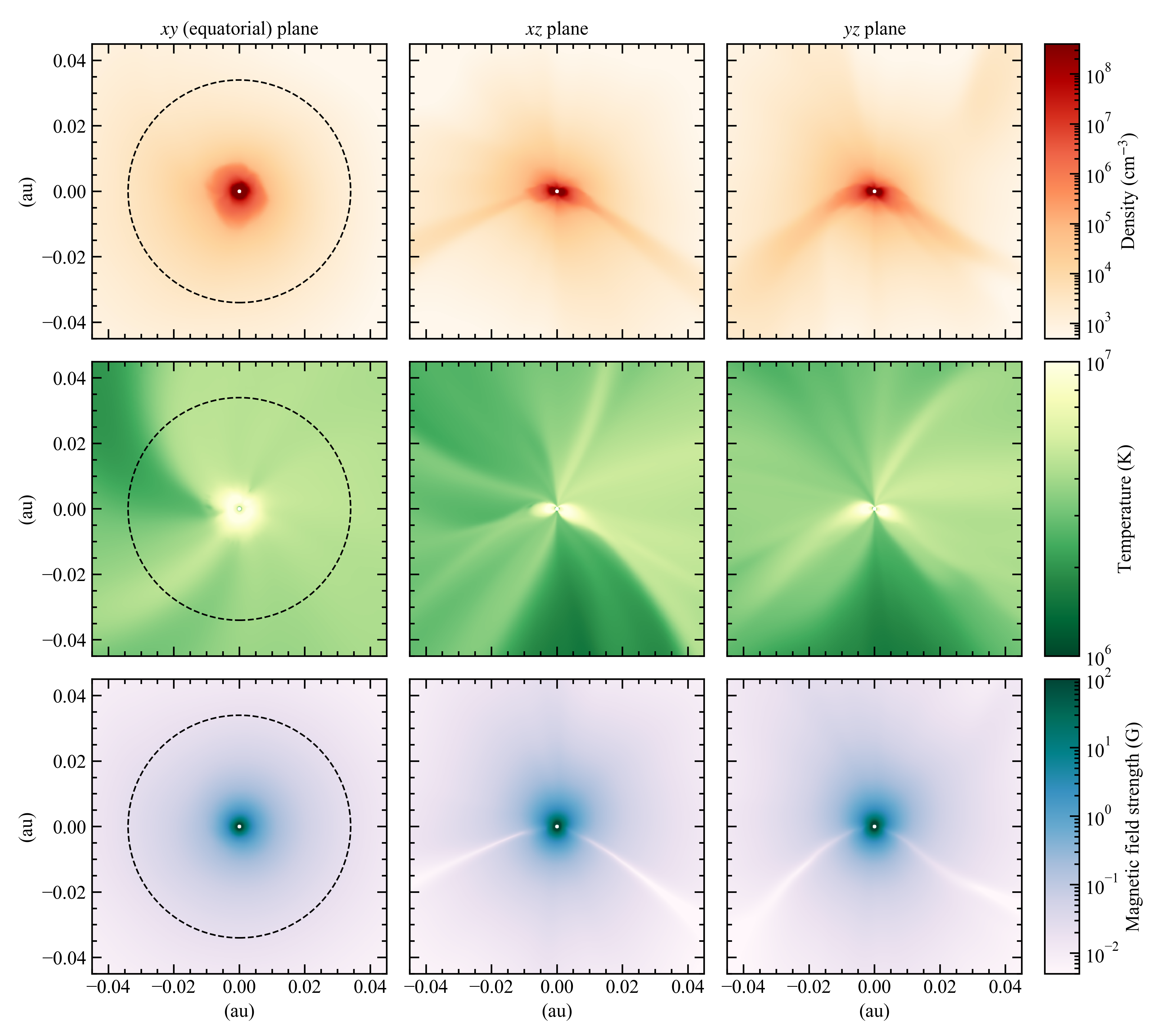}
\caption{Profiles of the stellar wind density (top), electron temperature (middle), and magnetic field strength (bottom) of WX UMa, in the \textit{xy}, \textit{xz}, and \textit{yz} planes (left to right). In the equatorial (\textit{xy}) plane, we show the orbital distance (0.034~au) where a potential Neptune-sized planet can reproduce the observed radio emission from the star at 144~MHz through star-planet interactions. The \textit{z}-axis is the rotation axis of the star.}
\label{fig:wind profiles}
\end{figure*}

In Figure~\ref{fig:wind profiles} we present profiles of the stellar wind density, electron temperature, and magnetic field strength for each of the Cartesian planes in our model.

\section{Power generated in sub-Alfv\'enic star-planet interactions}
\label{sec:planet power derivation}

The power generated via the interaction between the stellar magnetic field and a planet in a sub-Alfv\'enic orbit is given by Equation~(55) of \citet{saur13} in SI units:
\begin{equation}
P = \frac{2 \pi}{\mu_0} {R_\textrm{m}}^2 (\alpha M_A B_\textrm{w} \cos \Theta)^2 u_\textrm{A} .
\label{eq:planet power saur}
\end{equation}
Here $\mu_0$ is the permeability of free space, $\alpha$ is the interaction strength, which is $\approx1$ for M-dwarfs \citep{turnpenney18}, $M_\textrm{A} = \Delta u / u_\textrm{A}$ is the Alfv\'enic Mach number, and $\Theta$ is the angle between vector $\Delta\boldsymbol{u}$ and the perpendicular component of the vector $\boldsymbol{B}_\textrm{w}$. We write our Equation~\ref{eq:planet power} in terms of $\theta$, which is the angle between the vectors $\Delta\boldsymbol{u}$ and $\boldsymbol{B}_\textrm{w}$. Comparing to \citet{saur13}, $\theta = \pi/2 - \Theta$, and so in Equation~\ref{eq:planet power saur} we have $\cos \Theta \equiv \sin \theta$. Rewriting Equation~\ref{eq:planet power saur} in CGS units, we then have
\begin{equation}
P = \frac{1}{2} R^2 (M_\textrm{A} B_\textrm{w} \sin \theta)^2 u_\textrm{A} ,
\end{equation}
and finally, expanding the factor ${M_\textrm{A}}^2 / u_\textrm{A}$ gives
\begin{equation}
P = \pi^{1/2} R^2 B_\textrm{w} {\rho_\textrm{w}}^{1/2} \Delta u^2 \sin^2\theta .
\end{equation}


\section{Dynamic radio spectrum of second harmonic ECMI}

\begin{figure*}
\centering
\includegraphics[width = \textwidth]{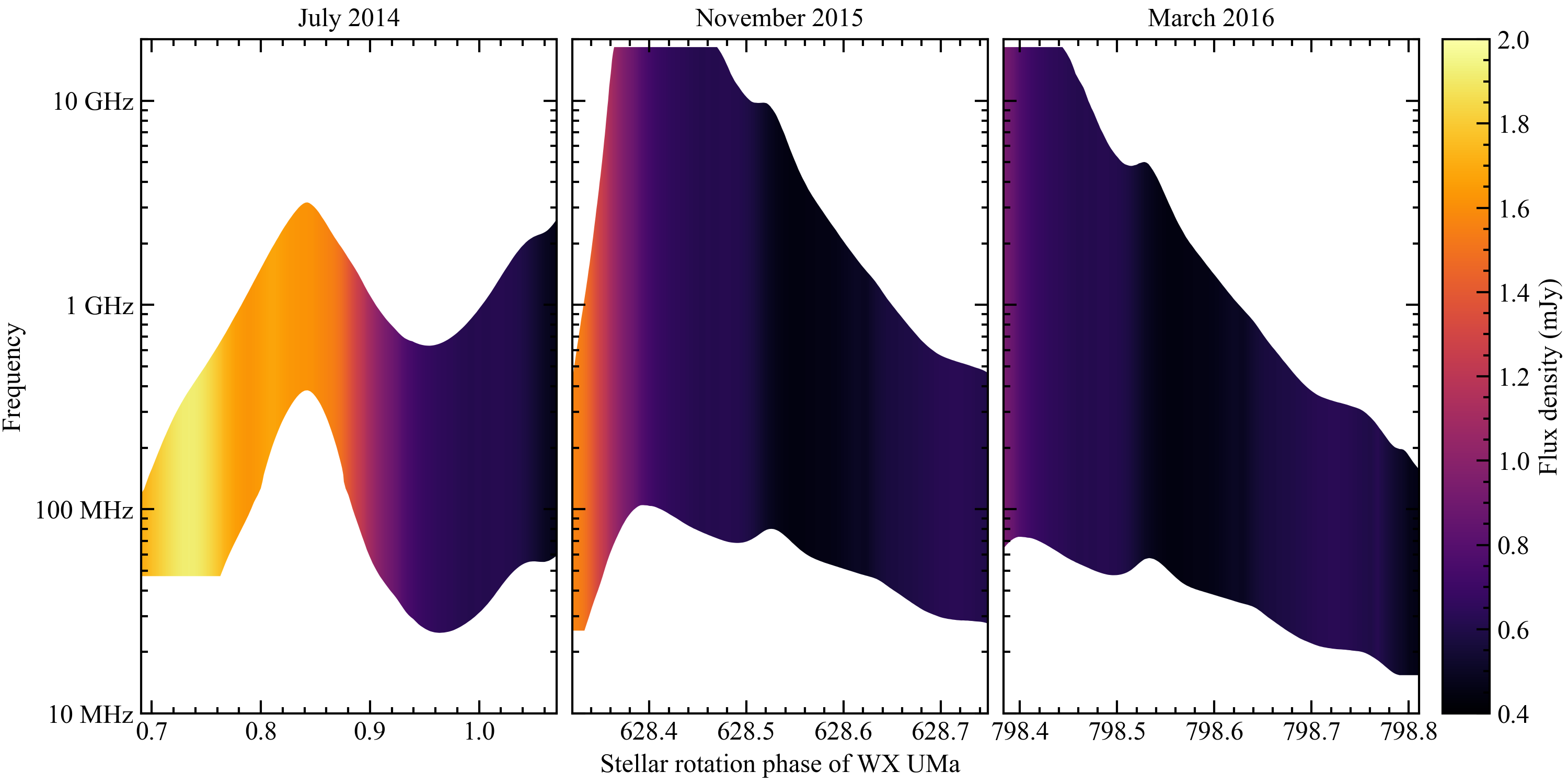}
\caption{Same as Figure~\ref{fig:dynamic spectrum - fundamental}, but for second harmonic ECMI emission.}
\label{fig:dynamic spectrum - harmonic}
\end{figure*}

Here, we show the dynamic radio spectrum induced by the planet that best-reproduces the observed emission from WX~UMa at 144~MHz, assuming it occurs via second harmonic ECMI. Emission can be generated from 20 MHz to 20 GHz for the field lines connecting to the orbit at $60.6~R_\star$, twice that of the dynamic spectrum for fundamental emission shown in Figure~\ref{fig:dynamic spectrum - fundamental}. The slight differences in morphology between the fundamental and harmonic dynamic spectra are due to the different locations of the emitting points. The second harmonic dynamic radio spectrum is shown in Figure~\ref{fig:dynamic spectrum - harmonic}.

\section{Propagation effects at 144 MHz}
\label{sec:propagation quantities}

\begin{figure*}
\centering
\includegraphics[width = \textwidth]{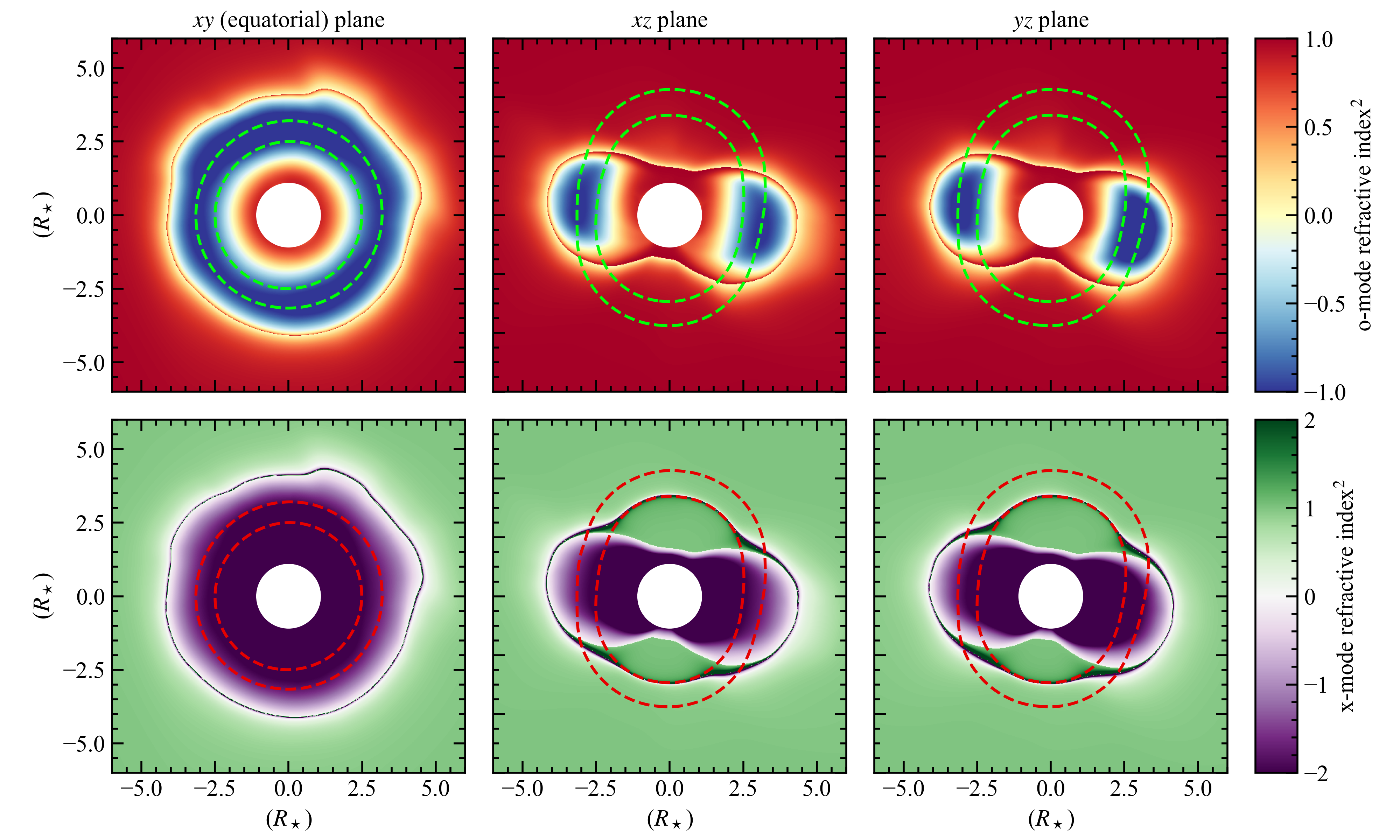}
\caption{Square of the magnetoionic refractive indices of \textit{o} (top) and \textit{x}-mode emission (bottom) at 144~MHz of the wind of WX~UMa. For a negative squared refractive index, the refractive index is imaginary, and the emission can no longer escape without being absorbed \citep{leto19}. The two dashed contours in each panel show the region where fundamental (inner) and second harmonic (outer) ECMI can be generated at 144~MHz.}
\label{fig:refractive indices}
\end{figure*}

As laid out in Section~\ref{sec:model limitations}, there are propagation effects that are not taken into account in our model, which may significantly alter the resulting observed emission from WX~UMa. Here, we compute both the refractive indices of \textit{o} and \textit{x}-mode emission, as well as the optical depth of the stellar wind at 144~MHz. The square of the refractive index for \textit{o/x}-mode emission is \citep{melrose82}:
\begin{equation}
{n_{\textit{o,x}}}^2 = 1 - \frac{X T_{\textit{o,x}}}{T_{\textit{o,x}} - Y \cos \alpha},
\end{equation}
where 
\begin{equation}
T_\textit{o} = - {T_\textit{x}}^{-1} = - (Z^2 + 1) ^ {1/2} - Z ,
\end{equation}
\begin{equation}
Z = \frac{Y\sin^2\alpha}{2 (1 - X) \cos\alpha} ,
\end{equation}
\begin{equation}
Y = \frac{\nu_\textrm{c}}{\nu} ,
\end{equation}
and
\begin{equation}
X = \Big( \frac{\nu_\textrm{p}}{\nu} \Big)^2 .
\end{equation}
In \citet{melrose82}, the above expressions are given as functions of the angle between the beam direction and the magnetic field. While in our analysis we consider beam angles for $\alpha\pm\Delta\alpha/2$, for brevity here we just compute the refractive indices for $\alpha$. In the case where the refractive indices are real, the emission will deviate from its path as it propagates through the magnetosphere. However, if the refractive indices are imaginary, the emission will become attenuated \citep{leto19}. In Figure~\ref{fig:refractive indices} we show the values of $n_{\textit{o,x}}^2$ in the three Cartesian planes of the stellar wind. In general, \textit{x}-mode emission is more likely to be absorbed as it propagates outwards, and naturally being generated further from the star, second harmonic ECMI can more readily escape. Regardless, Figure~\ref{fig:refractive indices} illustrates that ECMI is likely to deviate from a straight line as it propagates outwards, passing in and out of attenuating regions in the process. Models which self-consistently account for these effects \citep[e.g.][]{lo12, leto16, das20} coupled with the planet-induced and reconnection models presented here are certainly worth exploring in the future. 

Thermal free-free absorption may also attenuate the emission as it escapes the system. Here, we compute the free-free absorption of the stellar wind in the same manner as \citet{ofionnagain19} and \citet{kavanagh20}. The absorption coefficient for free-free emission in CGS units is \citep{rybicki86}:
\begin{equation}
\alpha_\nu = 3.692\times10^8 Z^2 T^{-1/2} \nu^{-3} n_\textrm{e} n_\textrm{i} g (1 - e^{-h\nu/kT}) .
\end{equation}
Here, $Z$ is the ionisation fraction of the plasma, $n_\textrm{e}$ and $n_\textrm{i}$ are the electron and ion number densities respectively, and $g$ is the Gaunt factor, which at radio frequencies is \citep{cox00}:
\begin{equation}
g = 10.6 + 1.90\log_{10}(T) - 1.26\log_{10}(Z\nu) .
\end{equation}
As the wind is composed of fully ionised hydrogen, with temperatures of $\gtrsim1$~MK (Figure~\ref{fig:wind profiles}), we set $Z=1$, and $n_\textrm{e} = n_\textrm{i} = n$. The optical depth at each emitting point is the integral of the absorption coefficient along the line of sight $x'$:
\begin{equation}
\tau_\nu = \int_{-\infty}^{{x_0}'} \alpha_\nu dx',
\end{equation}
where ${x_0}'$ is the coordinate of the emitting point along the line of sight. Numerically, we replace the lower limit of the integral with max distance at which the wind is dense enough to increase the optical depth significantly (at around $10~R_\star$). The optical depth then allows us to compute the factor the escaping emission is attenuated by, which is $e^{-\tau_\nu}$ \citep{kavanagh20}.

\begin{figure*}
\centering
\includegraphics[width = \textwidth]{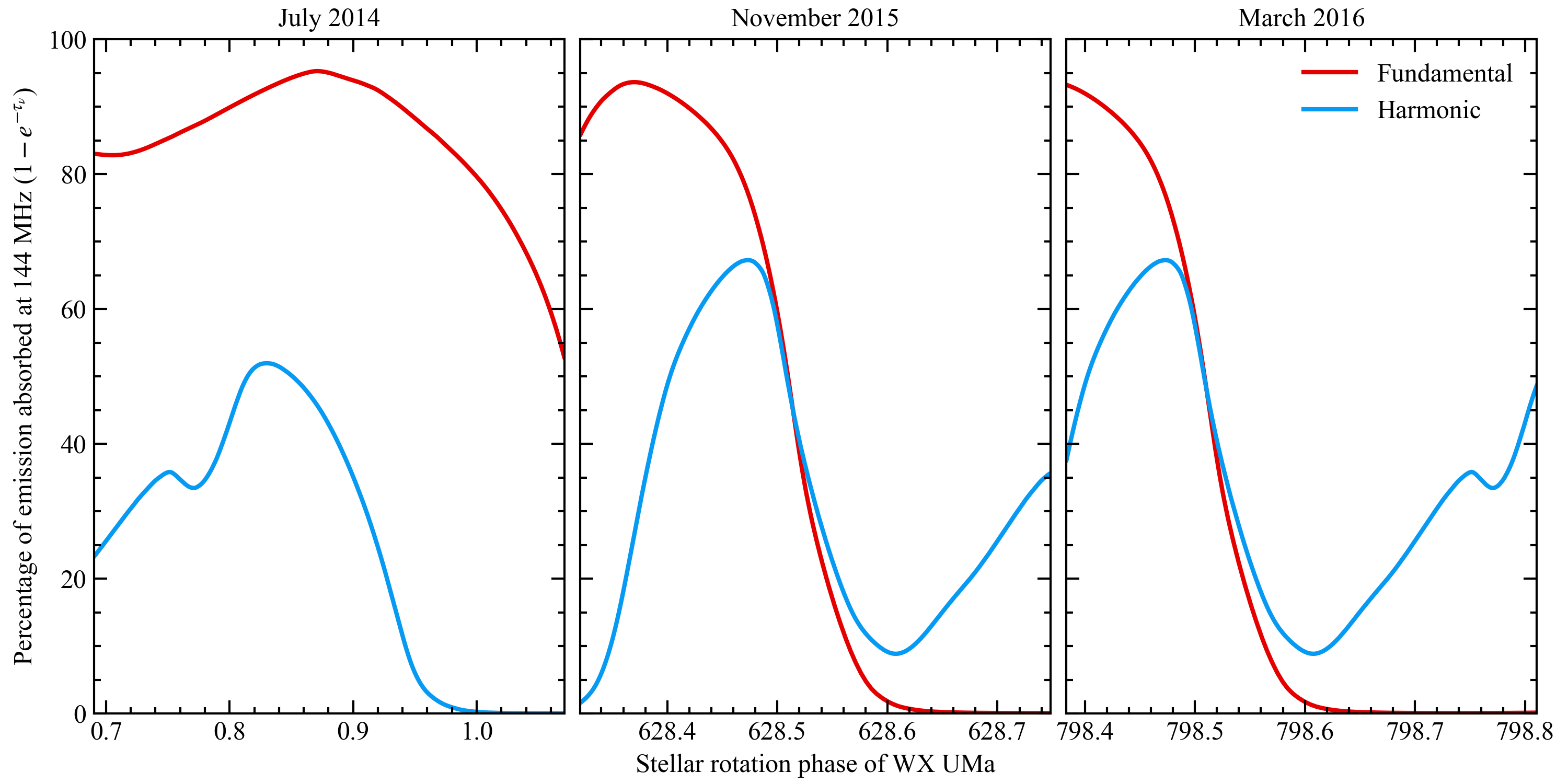}
\caption{The fraction of radio emission induced in WX~UMa by a planet at $\sim0.034$~au that is absorbed due to free-free processes. Fundamental ECMI is more readily absorbed, as the points generating emission at this mode are more deeply embedded in the stellar wind, resulting in larger optical depths.}
\label{fig:optical depth}
\end{figure*}

We compute the optical depth at the points along the stellar magnetic field lines connecting to the orbit of the potential planet identified in Section~\ref{sec:best fit} that best-reproduces the LOFAR observations, accounting for the stellar inclination and rotation. The results are shown in Figure~\ref{fig:optical depth}. We find a varying level of attenuation for emission during the three epochs, with fundamental ECMI being absorbed on average by 51\%. Again, second harmonic ECMI is less attenuated, with 30\% being absorbed on average. This is due to the emission being generated further from the star, which results in lower optical depths along the line of sight. These results clearly illustrate that accounting for free-free absorption, alongside magnetoionic refraction/absorption, should be considered in the model in a self-consistent way for future applications.

\bsp
\label{lastpage}
\end{document}